%% file: templateArxiv.tex
\documentclass{article}

\usepackage{PRIMEarxiv}

\usepackage[utf8]{inputenc} % allow utf-8 input
\usepackage[T1]{fontenc}    % use 8-bit T1 fonts
\usepackage{hyperref}       % hyperlinks
\usepackage{url}            % simple URL typesetting
\usepackage{booktabs}       % professional-quality tables
\usepackage{amsfonts}       % blackboard math symbols
\usepackage{nicefrac}       % compact symbols for 1/2, etc.
\usepackage{microtype}      % microtypography
\usepackage{lipsum}
\usepackage{fancyhdr}       % header
\usepackage{graphicx}       % graphics
\graphicspath{{media/}}     % organize your images and other figures under media/ folder

\usepackage{amsmath}
\usepackage{algorithm}
\usepackage{algpseudocode}
\usepackage{threeparttable}
\usepackage{multirow}

\title{A Bayesian two-step multiple imputation approach based on mixed models for the missing in EMA data} 

\author{
  Yiheng Wei \\
  Committee on Computational and Applied Mathematics \\
  The University of Chicago \\
  Chicago, IL 60637\\
  \texttt{yiheng@uchicago.edu} \\
   \And
  Donald Hedeker \\
  Department of Public Health Sciences \\
  The University of Chicago \\
  Chicago, IL 60637\\
  \texttt{hedeker@uchicago.edu} \\
}

\begin{document}
\maketitle

\begin{abstract}
Ecological Momentary Assessments (EMA) capture real-time thoughts and behaviors in natural settings, producing rich longitudinal data for statistical and physiological analyses. However, the robustness of these analyses can be compromised by the large amount of missing in EMA data sets. To address this, multiple imputation, a method that replaces missing values with several plausible alternatives, has become increasingly popular. In this paper, we introduce a two-step Bayesian multiple imputation framework which leverages the configuration of mixed models. We adopt the Random Intercept Linear Mixed model, the Mixed-effect Location Scale model which accounts for subject variance influenced by covariates and random effects, and the Shared Parameter Location Scale Mixed Effect model which links the missing data to the response variable through a random intercept logistic model, to complete the posterior distribution within the framework. In the simulation study and an application on data from a study on caregivers of dementia patients, we further adapt this two-step Bayesian multiple imputation strategy to handle simultaneous missing variables in EMA data sets and compare the effectiveness of multiple imputations across different mixed models. The analyses highlight the advantages of multiple imputations over single imputations. Furthermore, we propose two pivotal considerations in selecting the optimal mixed model for the two-step imputation: the influence of covariates as well as random effects on the within-variance, and the nature of missing data in relation to the response variable.
\end{abstract}

% keywords can be removed
\keywords{Ecological momentary assessments \and Longitudinal data \and Mixed model \and Shared parameter model \and Missing not at Random \and Caregiver \and Dementia}

\section{Introduction}\label{introduction}
Ecological Momentary Assessment (EMA) involves the repetitive sampling of subjects' current behaviors and experiences in their natural environments, which aims to assess particular events in subjects’ lives or assess subjects at periodic intervals. The methodology is to do random time sampling, which employs a variety of technologies, ranging from traditional written diaries and telephones to more modern electronic diaries and physiological sensors. The use of EMA allows researchers to compile longitudinal datasets for different subjects, with each subject contributing numerous repeated observations over varying time spans, as designed by the experiment \cite{shiffman2008ecological}. However, it is usually the case that some observations are missing due to non-responses from the subjects, leading to an incomplete dataset.

In longitudinal analysis, there are typically three types of missing data mechanisms: the absence of observations is purely random and unrelated to any inferences we aim to draw about the intervention effect, which is referred to as Missing Completely at Random (MCAR); the missing mechanism has an association between the chance of missing and observed data, which is Missing at Random (MAR); the missing is also determined by the unobserved values of observations, which is Missing not at Random (MNAR) \cite{little2019statistical}. A common and straightforward method for dealing with missing data is to exclude observations with missing records and work only with complete observations. If the missingness mechanism is MCAR, a complete case analysis can be sensible, although it may well not use all the available information in the data \cite{carpenter2007missing}. However, in practice, most missing data scenarios do not adhere to the MCAR assumption. When data is not missing completely at random, conducting a complete case analysis can lead to a different inference result \cite{burton2007cost}. Consequently, imputing the missing values becomes a crucial task to maximize the utilization of available data and ensure the validity of statistical analyses.

To address missing data in longitudinal data sets, various methods have been developed, such as the pattern mixture models \cite{little1993pattern, hedeker1997application}, the selection models \cite{diggle1994informative}, and the shared parameter models \cite{vonesh2006shared}. Apart from these methods, multiple imputation has emerged as an effective approach, which is to replace missing by generating $m$ different acceptable values representing a distribution of possibilities \cite{rubin2004multiple}. The theorems and methodology of multiple imputations have been well-established by Schafer \cite{schafer1999multiple}. In our study, we employ the two-step Bayes approach by Schafer, which can be derived explicitly through formulas, for multiple imputations: by assuming a posterior distribution for the data $y$ given a parameter set $\theta$, we firstly derive the posterior distribution of $\theta$ based on the observed data and draw $m$ different $\theta$ values from this posterior distribution, then we draw $m$ imputed values from the posterior distribution of $y$, conditioning on each of the $\theta$ values. 

Considering the statistical structure of EMA data, one common approach to constructing the posterior distribution of the data $y$ on a given parameter set is through the utilization of a random effect model \cite{laird1982random}, which is also known as the linear mixed model. In order to accommodate the more intricate patterns for the within variance, Hedeker et al \cite{hedeker2008application} expanded upon this approach by incorporating log-linear models for both within-subject (WS) and between-subject (BS) variances. This extension allows for the potential influence of covariates on both sources of variation and permits the inclusion of a random subject effect in the WS variance specification. This model was successfully applied in an EMA study on adolescent smoking to characterize the changes in mood variation. However, thus far, these models are typically constructed using datasets with missingness and do not explicitly take into account the missing data patterns. Building upon Cursio et al's work \cite{cursio2019latent}, who proposed a joint model of the intensive longitudinal data and the missingness of the data, Lin et al\cite{lin2018shared} proposed a shared parameter modeling approach that links the primary longitudinal outcome with potentially informative missingness by introducing a random effect on the missing pattern for each subject. Lin also proposed the estimation approach via a fully Bayesian approach using the Markov Chain Monte Carlo (MCMC) method. By establishing a connection between the observed data and the potential missingness, this model effectively represents an MNAR setting, which is more general in real-world scenarios.

In this paper, we will establish the two-step Bayes approach for multiple imputations based on different mixed models, and compare the imputation performance under different aspects and criteria. In terms of the organization of this paper, the example EMA data set is described in Section \ref{motivating example}, and the statistical analysis on the models, imputation method, and estimation approach is provided in Section \ref{statistical analysis}. In Section \ref{simulation study} and \ref{example}, we will base the simulated EMA data and the real data set to analyze the performance of our multiple imputations. The conclusion and discussion are provided in Section \ref{discussion}.

\section{Motivating example}\label{motivating example}
This research is motivated by Van Knippenberg's study on the characteristics that modify emotional stress reactivity in the caregivers of dementia patients \cite{van2018emotional}. The study contains a total of 30 subjects, spanning a duration of 6 days. The data set comprises several demographic characteristics of the caregivers, including gender, age, and education. It also contains some demographic features of the patients, including the duration of dementia in years, and their scores on the Clinical Dementia Rating (CDR) scale. During the experiment, participating caregivers were provided with an EMA device known as the 'PsyMate' to collect the data in their daily lives. Within each day of the study, participants experienced ten random beeps at different, unpredictable moments, scheduled between 7:30 AM and 10:30 PM. After each beep, participants were required to assess their current stress states and Positive Affect (PA) states rated on a 7-point bipolar Likert scale, where 1 indicated very low and 7 indicated very high. In this study, we will consider both stress and PA as continuous variables to conform to the commonly employed normal distribution. Table \ref{Demographic and clinical variables of the caregivers and care recipients} presents descriptive statistics on the demographic and clinical variables of the 30 participating caregivers and their care recipients.

\begin{table}[ht]
    \centering
\caption{Demographic and clinical variables of the caregivers and care recipients}
\label{Demographic and clinical variables of the caregivers and care recipients}
\begin{tabular}{l|l}
\hline
\textbf{Variable}          & \textbf{Value} \\ \hline
Gender $(n, \%)$                     &                                           \\ \hline
\quad Female                    & 18 (60.00)                                 \\ \hline
\quad Male                  & 12 (40.00)                                 \\ \hline
Age $(M, SD, range)$                        & 69.87 $\pm$ 5.82 (57.00-80.00)                              \\ \hline
Level of education $(n, \%)$         &                                           \\ \hline
\quad 1: Low                     & 13 (43.33)                                 \\ \hline
\quad 2: Middle                  & 8 (26.67)                                  \\ \hline
\quad 3: High                    & 9 (30.00)                                  \\ \hline
Dementia duration in years $(M, SD, range)$ & 6.00 $\pm$ 3.84 (1.00-15.00)                               \\ \hline
CDR $(n, \%)$                        &                                           \\ \hline
\quad 1: Low                     & 22 (73.33)                                 \\ \hline
\quad 2: Middle                  & 7 (23.33)                                  \\ \hline
\quad 3: High                    & 1 (3.33)                                   \\ \hline
PA $(M, SD, range)$            & 5.09 $\pm$ 1.22 (1.00-7.00)                               \\ \hline
Stress $(M, SD, range)$                    & 2.71 $\pm$ 1.33 (1.00-7.00)                               \\ \hline
\end{tabular}
\end{table}

In the data set, we have information available for the demographic features, but intermittent missingness occurred for stress and PA when individuals did not respond to the scheduled beeps. Consequently, we will propose methods to impute these missing values. The overall percent of missing responses in the study was 19.56\%, although it varied across different time periods and among subjects. Specifically, the percent of missing 
responses for each of the six days of the study were 16.67\%, 21.00\%, 18.33\%, 19.33\%, 23.67\%, and 18.33\%, and the percent of missing responses for each of the ten scheduled beeps were 42.22\%, 20.56\%, 20.00\%, 13.89\%, 18.33\%, 21.67\%, 17.22\%, 16.67\%, 13.89\%, and 11.11\%. These results reveal a trend of lower missingness on the first day of the study, and higher missingness during the earlier prompts. Also, among the 30 subjects, the highest percent of missing data observed was 38.33\%, while the lowest was only 1.67\%, with a standard deviation between subjects as 9.30\%. This variability in missingness across subjects underscores the significance of modeling different missing patterns for different individuals.

\section{Statistical analysis}\label{statistical analysis}

\subsection{Models}
Let $y_{ij}$ represent the outcome for subject $i$ at occasion $j$, and $x_{ij}$ represent the $p \times 1$ covariate vector for subject $i$ at occasion $j$, where $i = 1, 2, \dots, n$ and $j = 1, 2, \dots, n_i$. In this two-level modeling framework, we are dealing with observations from various subjects, and within each subject, we have observations across different occasions. To capture the interactions between subject-level and occasion-level variables, we will begin with the Random Intercept Linear Mixed (RILM) model \cite{laird1982random}:

\begin{align}
&y_{ij} \mid x_{ij}, v_{1, i} \sim \mathcal{N}
\left(
\beta_0 + x^\top_{ij} \beta + v_{1, i} , \; e^{\alpha_0}
\right)
\\
&v_{1, i} \sim \mathcal{N}
\left(
0 , \sigma^2_{v_{1}}
\right)
,
\end{align}
where $\beta_0$ is the coefficient for intercept, $\beta$ is the $p \times 1$ vector of coefficients, and $v_{1, i}$ is what we call the random location effect for the subject-level observations, which follows a normal distribution with zero mean and fixed variance $\sigma^2_{v_{1}}$. To align with the expressions of the following models, we use an exponential expression, $e^{\alpha_0}$, to model the fixed WS variance.

To allow covariates to influence the WS variance, we can expand the model by including the covariates in the modeling of within-subject variance and also a random scale effect which can be correlated with the random location effect. This model is called the Mixed-effects Location Scale (MELS) model \cite{hedeker2008application}:

\begin{align}
&y_{ij} \mid x_{ij}, v_{1, i}, v_{2, i} \sim \mathcal{N}
\left(
\beta_0 + x^\top_{ij} \beta + v_{1, i} , \; e^{\alpha_0 + x_{ij}^\top \alpha + v_{2, i}}
\right)\\
&\begin{pmatrix}
    v_{1, i} \\ v_{2, i}
\end{pmatrix}
\sim
\mathcal{N}
\left(
\begin{pmatrix}
    0 \\ 0
\end{pmatrix}
,
\begin{pmatrix}
    \sigma_{v_{1}}^2 & \rho_{v_{1}, v_{2}}\sigma_{v_{1}}\sigma_{v_{2}}\\
    \rho_{v_{1}, v_{2}}\sigma_{v_{1}}\sigma_{v_{2}} & \sigma_{v_{2}}^2\\
\end{pmatrix}
\right)
,
\end{align}
where $\alpha_0$ is the coefficient for the intercept and $\alpha$ is the $p \times 1$ vector of coefficients in the model of $y_{ij}$'s conditional variance. $v_{2, i}$ is the random scale effect for different subjects, which follows a normal distribution with zero mean and a fixed variance $\sigma^2_{v_{2}}$. $\rho_{v_{1}, v_{2}}$ is the correlation between the random location effect $v_{1, i}$ and random scale effect $v_{2, i}$.

In EMA data, missing can occur and exhibit diverse patterns for different subjects and occasions. In Section \ref{motivating example}, we have shown that the missing probability is different across subjects, days, and beeps, underscoring the need for a model that accounts for the missing probability across subjects and occasions. Furthermore, the mean and variance of $y_{ij}$ can also be related to the missingness. For instance, individuals with tight schedules may experience higher stress levels and be less likely to respond to the scheduled beeps. Hence, it is reasonable to include some random effects representing the missing in the modeling of both the mean and variance of $y_{ij}$. Continuing on the MELS's setting, we employ the Shared Parameter Location Scale Mixed Effect (SPLSME) model \cite{lin2018shared} here. This model divides the random location and scale effects into one missing-decided part and one missing-orthogonal part:

\begin{align}
&v_{1, i} = \gamma \lambda_i +\eta_{1, i} \label{de1}\\
&v_{2, i} = \delta \lambda_i +\eta_{2, i}\label{de2},
\end{align}
where $\lambda_i$ denotes the random missing effect extracted from the model of missing:

\begin{align}
m_{ij} \mid t_{ij}, \lambda_i \sim \mathcal{B}
\left(
L\left(\tau_0 + t_{ij}^\top \tau + \lambda_i
\right)\right)
\label{model_miss} .
\end{align}

In the Formulas \ref{de1} and \ref{de2}, $\gamma$ and $\delta$ are two parameters that indicate how much the random missing effect $\lambda_i$ will determine the random location effect $v_{1. i}$ and the random scale effect $v_{2, i}$, respectively. $\eta_{1, i}$ and $\eta_{2, i}$ are two error terms that absorb the residual variance $\sigma^2_{\eta_1}$ and $\sigma^2_{\eta_2}$ which are orthogonal to $\lambda_i$. We call these two parameters the residual random location effect and the residual random scale effect. Since we allow the correlation between the random location effect $v_{1, i}$ and random scale effect $v_{2, i}$, there will also exist the correlation $\rho_{\eta_1, \eta_2}$ between $\eta_{1. i}$ and $\eta_{2, i}$. In the Formula \ref{model_miss},$L$ denotes the logistic function, and $m_{ij}$ denotes the missing indicator for subject $i$ at occasion $j$, which follows a Bernoulli distribution and is determined by some covariates and subject $i$'s random missing effect. $\tau_0$ is the intercept. $t_{ij}$ denotes the $q \times 1$ vector of covariates, and $\tau$ is the corresponding $q \times 1$ vector of coefficients. 

In summary, SPLSME is structured as follows:

\begin{align}
&y_{ij} \mid x_{ij}, \eta_{1, i}, \eta_{2, i}, \lambda_i \sim \mathcal{N}
\left(
\beta_0 + x^\top_{ij} \beta + \eta_{1, i} + \gamma \lambda_i ,
e^{\alpha_0 + x_{ij}^\top \alpha + \eta_{2, i} + \delta \lambda_i}
\right)
\\
&m_{ij} \mid t_{ij}, \lambda_i \sim \mathcal{B}
\left(
L\left(\tau_0 + t_{ij}^\top \tau + \lambda_i
\right)\right)
\\
&\begin{pmatrix}
    \eta_{1, i} \\ \eta_{2, i} \\ \lambda_i
\end{pmatrix}
\sim
\mathcal{N}
\left(
\begin{pmatrix}
    0 \\ 0 \\ 0
\end{pmatrix}
,
\begin{pmatrix}
    \sigma_{\eta_{1}}^2 & \rho_{\eta_{1}, \eta_{2}}\sigma_{\eta_{1}}\sigma_{\eta_{2}} & 0\\
    \rho_{\eta_{1}, \eta_{2}}\sigma_{\eta_{1}}\sigma_{\eta_{2}} & \sigma_{\eta_{2}}^2 & 0\\
    0 & 0 & \sigma^2_{\lambda}
\end{pmatrix}
\right) 
.
\end{align}

\subsection{Multiple imputation}

Our imputation will follow a two-step parametric Bayesian setting  \cite{schafer1999multiple}. Denote $y_{obs}$ as the observed data and $y_{miss}$ as the missing data, $\theta$ is a parameter set for $y$'s distribution. 
 Because

\begin{align}
    P\left(y_{miss} \mid y_{obs}\right) = \int P\left(y_{miss} \mid y_{obs}, \theta\right) P\left(\theta \mid y_{obs} \right) d\theta ,
\end{align}
an imputation for $y_{miss}$ can be created by first sampling

\begin{align}
    \hat\theta \sim P\left(\theta \mid y_{obs} \right) ,
\end{align}
and then imputing the $y_{miss}$ by sampling

\begin{align}
    \hat{y}_{miss} \sim P\left(y_{miss} \mid y_{obs}, \hat\theta\right).
\end{align}

\noindent
By repeating the sampling $m$ times, we can obtain $m$ estimated results of the combination $\{\hat\theta^{(k)}, \hat{y}^{(k)}_{miss}\}_{k = 1, \dots, m}$. The estimation of ${y_{miss}}$ can be calculated by $\frac1m \sum_{k = 1}^m {\hat{y}^{(k)}_{miss}}$.

Specifying the conditional distribution of $y$ and the corresponding parameter set $\theta$ is required before sampling. In this study, we will adopt the RILM, MELS, and SPLSME models discussed previously.  
The equations and discussion above provide a general overview of the two-step multiple imputation approach and the details for each of the three models can be found in Appendix \ref{appendix1}. 

\subsection{Parameter estimation}

The methods for estimating the parameters $\theta$ differ among these three models. For RILM, numerous existing packages are available that can yield results efficiently. In this study, we utilize the linear mixed-effects models using 'Eigen' and the S4 (lme4)\cite{lme4} package in R which employs maximum likelihood estimation. For the other two models, we adopt the Metropolis-Hastings algorithm \cite{metropolis1953equation, hastings1970monte} to avoid the multi-dimensional integration in the maximum likelihood estimation method. This algorithm is based on the Markov Chain Monte Carlo (MCMC) theorems, and revised by Lin et al \cite{lin2018shared} to accommodate the model specifications of MELS and SPLSME: 

\begin{algorithm}
\caption{\enskip Metropolis-Hastings sampling algorithm}\label{alg1}
\begin{algorithmic}
\State Input initial point $\theta^{(0)}$.
\For {$t = 1, 2, \dots, T$}
\State Sample $\theta ^* \sim  q(\theta ^{(t-1)}, \theta ^*)$.
\State Update 
\begin{align}
\theta^{(t)}  = 
\begin{cases}
\theta ^* \quad &\text{w.p. } a(\theta ^{(t-1)}, \theta ^*) \\
\theta^{(t-1)}  \quad &\text{w.p. } 1 - a(\theta ^{(t-1)}, \theta ^*)
\end{cases} .
\end{align}
\EndFor
\State Output the conditional distribution $P\left(\theta \mid y_{obs} \right)$.
\end{algorithmic}
\end{algorithm}

The Markov kernel $q(\theta^{(t-1)}, \theta^* )$ should be set before running the algorithm, and the acceptance probability in this algorithm is
\begin{align}
    a(\theta ^{(t-1)}, \theta ^*) = \min\left(1, \frac{P(\theta ^* \mid y_{obs} ) 
    q(\theta ^*, \theta ^{(t-1)})}
    {P(\theta ^{(t-1)} \mid y_{obs} ) 
    q(\theta ^{(t-1)}, \theta^*)}\right) .
\end{align}

Through the Bayesian formula, the full conditional distribution in the algorithm can be derived by

\begin{align}
    P \left(\theta \mid  y_{obs} \right) \propto L \left(y_{obs} \mid \theta \right) P\left(\theta \right) ,
\end{align}
where $P\left (\theta\right)$ denotes the priors, and $L \left(y_{obs} \mid \theta\right)$ denotes the likelihood. 

By continuing sampling and updating each parameter, the algorithm eventually converges to the invariant distribution of this Markov chain, which is exactly the posterior distribution $P\left(\theta \mid y_{obs} \right)$. In our study, we adopt Rstan to carry out this procedure, a tool that provides full Bayesian inference for continuous-variable models through Markov Chain Monte Carlo methods such as the No-U-Turn sampler, an adaptive form of Hamiltonian Monte Carlo sampling \cite{rstan}. This subsection provides a generalized overview of the estimation methods used, and for more details of Algorithm \ref{alg1} related to each specific model, please refer to Appendix \ref{appendix2}.

\section{Simulation study}\label{simulation study}
To conduct a performance comparison between single imputation and multiple imputation methods, as well as between imputation methods based on different modeling settings, we simulated a dataset including two covariates, denoted as ${x_1}_{ij}$ and ${x_2}_i$, along with the response variable $y_{ij}$. ${x_1}_{ij}$ and $y_{ij}$ are continuous variables that change over subjects and occasions, similar to the PA and stress variables in the caregivers' data set. ${x_2}_i$ is a subject-level variable resembling demographic features. 
Based on the caregivers' data set, the simulated data set contains 20 subjects, 5 days, and 6 beeps within a day. 

For some beeps, both the values of $y_{ij}$ and ${x_1}_{ij}$ are missing. We adopt the SPLSME, which is an MNAR model setting, to create the missing for $y_{ij}$ and ${x_1}_{ij}$. Usually, missing tends to increase with experiment days, and also during the early morning and late night. To mimic the missing pattern of the real data set, we introduce three variables: $t_{ij}$, ${b_1}_{ij}$, and ${b_6}_{ij}$. $t_{ij}$ takes on integers ranging from 1 to 5, reflecting the different days of data collection. ${b_1}_{ij}$ and ${b_6}_{ij}$ are binary variables that indicate whether the response corresponds to the first beep or the sixth beep, respectively. These beeps occur during the early morning and late night. Incorporating these variables into the simulation allows us to emulate the missing pattern observed in the actual data set, where certain responses are more likely to be missing during specific times of the day and across different days of data collection. What's more, to control the missing ratio to around 20\% in each simulation, we adjust the intercept for the missing model, which is denoted as $\tau_0$, to control the missing ratio. The simulated process can be summarized as the following four steps:

\begin{enumerate}
\item Simulate ${x_2}_i$ by
\begin{align*}
{x_2}_i \sim \mathcal{N} 
\left(\mu_{{x_2}}, \sigma^2_{x_2}\right) .
\end{align*}

\item Simulate the random missing effect $\lambda_i$ and missing indicator $m_{ij}$ by
\begin{align*}
&\lambda_i \sim \mathcal{N} \left(0, \sigma^2_{\lambda} \right) \\
&m_{ij} \sim \mathcal{B} \left(L\left(\tau_0 + \tau_1 t_{ij} + \tau_2 {b_1}_{ij} + \tau_3 {b_3}_{ij} + \lambda_i\right)\right) .
\end{align*}

\item Simulate ${x_1}_{ij}$ by
\begin{align*}
&\begin{pmatrix}
    \eta^{(1)}_{1, i} \\ \eta^{(1)}_{2, i} 
\end{pmatrix}
\sim
\mathcal{N}
\left(
\begin{pmatrix}
    0 \\ 0 
\end{pmatrix}
,
\begin{pmatrix}
    \sigma_{\eta_1^{(1)}}^2 & \rho_{\eta_1^{(1)}, \eta_2^{(1)}}\sigma_{\eta_1^{(1)}}\sigma_{\eta_2^{(1)}} \\
    \rho_{\eta_1^{(1)}, \eta_2^{(1)}}\sigma_{\eta_1^{(1)}}\sigma_{\eta_2^{(1)}} & \sigma_{\eta_2^{(1)}}^2
\end{pmatrix}
\right) \\
&{x_1}_{ij} \mid {x_2}_i, \eta^{(1)}_{1, i}, \eta^{(1)}_{2, i}, \lambda^{(1)}_i 
\sim 
\mathcal{N}
\left( \beta^{(1)}_0 + \beta_2^{(1)} {x_2}_i + \eta^{(1)}_{1, i} + \gamma^{(1)} \lambda_i
,
e^{\alpha^{(1)}_0 + \alpha_2^{(1)} {x_2}_i + \eta^{(1)}_{2, i} + \delta^{(1)} \lambda_i} \right) .
\end{align*}

\item Simulate $y_{ij}$ by

\begin{align*}
&\begin{pmatrix}
    \eta^{(2)}_{1, i} \\ \eta^{(2)}_{2, i} 
\end{pmatrix}
\sim
\mathcal{N}
\left(
\begin{pmatrix}
    0 \\ 0 
\end{pmatrix}
,
\begin{pmatrix}
    \sigma_{\eta_1^{(2)}}^2 & \rho_{\eta_1^{(2)}, \eta_2^{(2)}}\sigma_{\eta_1^{(2)}}\sigma_{\eta_2^{(2)}} \\
    \rho_{\eta_1^{(2)}, \eta_2^{(2)}}\sigma_{\eta_1^{(2)}}\sigma_{\eta_2^{(2)}} & \sigma_{\eta_2^{(2)}}^2
\end{pmatrix}
\right) \\
&y_{ij} \mid {x_1}_{ij}, {x_2}_i, \eta^{(2)}_{1, i}, \eta^{(2)}_{2, i}, \lambda^{(2)}_i
\sim 
\mathcal{N}
\left(\beta^{(2)}_0 + \beta_1^{(2)} {x_1}_{ij} + \beta_2^{(2)} {x_2}_i + \eta^{(2)}_{1, i} + \gamma^{(2)} \lambda_i
,
e^{\alpha^{(2)}_0 + \alpha_1^{(2)} {x_1}_{ij} + \alpha_2^{(2)} {x_2}_i + \eta^{(2)}_{2, i} + \delta^{(2)} \lambda_i}\right) .
\end{align*}
\end{enumerate}

For imputation, we first adopt the RILM, MELS, and SPLSME to impute ${x_1}_{ij}$ respectively, using all available covariates, including ${x_2}_{ij}$, $t_{ij}$, ${b_1}_{ij}$, and ${b_6}_{ij}$. Then, we also use the three models to impute $y_{ij}$ based on the imputed values of ${x_1}_{ij}$ and the other covariates. We repeat the imputation process 100 times both for single imputation and multiple imputation. The number of repeated sampling for multiple imputations is set to 10. We systematically modify some of the parameter values in the simulations, including $\alpha^{(2)}_0$, $\rho_{\eta_1^{(2)}, \eta_2^{(2)}}$, $\gamma^{(2)}$, $\delta^{(2)}$ that are used to generate $y_{ij}$.

The evaluation criteria we use include error, bias, and coverage rate. Error is calculated as the average of the squared differences between the real values of $y_{ij}$ and their imputed values $\hat y_{ij}$, which is $(\sum_{k = 1}^{100} (\frac{\sum_{i = 1}^{i = n} \frac{\sum_{j = 1}^{j = n_i}(\hat y_{ijk} - y_{ijk})^2}{n_i}}{n}))/100$. Bias is computed for each parameter as the deviation from the true value, which is $(\sum_{k = 1}^{100} (\hat\theta_k - \theta))/100$. Let $\hat\theta_k^l$ and $\hat\theta_k^u$ represent the lower and upper bounds, respectively, of the estimated 95\% confidence interval for the MLE method, or the credible interval for the MCMC estimation approach. The coverage rate is the percentage of the intervals that contain the true value, which is $\sum_{k=1}^{100} 1 _{\{\hat\theta_k^l < \theta < \hat\theta_k^u \}}$ mathematically.

\subsection{Modeling within-variance}\label{RILM and MELS}

Compared to RILM, both MELS and SPLSME allow the within variance to vary, depending on the covariates and random scale effect, and allow a correlation between the random location effect and random scale effect. To assess the individual contributions of introducing covariates and the random scale effect in the imputation process, we vary the values of $\alpha^{(2)}_0$ and $\rho_{\eta^{(2)}_{1}, \eta^{(2)}_{2}}$, and then calculate the corresponding errors, biases, and coverage rates.

\begin{table*}[t]
\small{
    \centering
    \input{error_12}}
\end{table*}

The errors are presented in Table \ref{error_12}. From Table \ref{error_12}, we can see that the errors of multiple imputations are always smaller than single imputations, no matter which model and parameter combination we use. Given this, our subsequent analysis will center on the influence of various values of parameters on the performance of multiple imputations. 

For the imputation of $x_1$, we consistently observed that SPLSME performs best, and then followed by MELS, and RILM is the worst by comparing the errors in Table \ref{error_12}. Since we only change the values of parameters used to simulate $y$ while keeping the parameters for simulating $x_1$ constant, the performance of imputing $x_1$ should be consistent.  By changing the parameters to simulate $x_1$, we can see a similar trend as changing the parameters to simulate $y$, thus we won't reanalyze this aspect here. However, one difference that should be considered is that when imputing $y$, we have an additional covariate, $x_1$, included in the model. Having fewer covariates in the model for $x_1$ results in more unexplained variance, and SPLSME might excel at capturing and explaining this unexplained variance by utilizing random effects and covariates effectively. 

To evaluate the performance of multiple imputations based on different models, for each method to impute $x_1$, Table \ref{error_12} provides the best model and worst models to impute $y$, and calculates the difference ratio between the best and worst models. When RILM is used to impute $x_1$, it's difficult to determine which model yields the smallest error for the imputation of $y$ since the errors from RILM, MELS, and SPLSME are closely comparable. This is because initiating with RILM for $x_1$ imputation introduces significant errors, leading to amplified errors regardless of the subsequent model chosen for $y$ imputation. However, when starting with MELS or SPLSME to impute $x_1$, it becomes evident that RILM consistently produces a larger error for the imputation of $y$ compared to MELS and SPLSME. 

When $\alpha^{(2)}_0$ is set to 0.00 and $\rho_{\eta_1^{(2)}, \eta_2^{(2)}}$ is set to -0.20, the lowest error is achieved when utilizing SPLSME for imputing ${x_1}$ and MELS for imputing $y$, both in the cases of single imputation and multiple imputation. Specifically, for single imputation, the smallest error is 8.57, and the largest error is 11.11. For multiple imputation, the smallest error is 8.00, and the largest error is 9.51. The difference ratios for the single imputation and multiple imputation are $\frac{11.11 - 8.57}{11.11} \times 100\% = 22.86\%$ and $\frac{9.51 - 8.00}{9.51} \times 100\% = 15.88\%$. This indicates that errors caused by insufficient model simplification can be mitigated by employing the multiple imputation method. 

\begin{table*}[t]
\small{
    \centering
    \input{bias_0_1_-0.2_1_1_-0.5_0.5_0.1_0.05_0.1}}
\end{table*}

Assessment of bias and coverage rates for $\alpha^{(2)}_0 = 0.00$ and $\rho_{\eta_1^{(2)}, \eta_2^{(2)}} = -0.20$ is provided in Table \ref{bias_0_1_-0.2_1_1_-0.5_0.5_0.1_0.05_0.1}. It is evident that the estimation obtained through MELS and SPLSME are generally good. The coverage rates for these consistently exceed 90\%, with most  achieving a coverage rate greater than or equal to 95\%. However, for RILM, the estimation of $\beta^{(2)}_1$ falls slightly short at 88\%. This discrepancy can be attributed to the fact that RILM, constrained by a fixed within-subject variance, cannot adequately account for changes in variance across occasions. Consequently, it transfers this variance change into ${x_1}$, the sole covariate that changes across occasions. Additionally, RILM displays notably poor performance in estimating $\alpha^{(2)}_0$ with a bias of -0.81 and a coverage rate of only 12\%. This arises from the fact that RILM relies solely on this parameter to mimic the variance pattern of $y_{ij}$ and attributes all the change of within-variance to $\alpha^{(2)}_0$, leading to a large bias for the estimation of $\alpha^{(2)}_0$.

As we increase the value of $\alpha^{(2)}_0$ from 0 to 1, 2, and 3, there is a corresponding increase in the variance of $y_{ij}$. This increase in variance introduces greater uncertainty in the imputation process, resulting in elevated imputation errors for both single and multiple imputation techniques. For example, in Table \ref{error_12}, the error of using RILM to impute both $x_1$ and $y$ for single imputation rises from 11.11 to 38.10 as $\alpha^{(2)}_0$ increases from 0 to 3. 

When $\alpha^{(2)}_0$ is small, the covariates and random scale effects exert a significant influence on the within-variance. However, as $\alpha^{(2)}_0$ becomes larger, the intercept value becomes dominant, reducing the model to an approximately RILM model. Calculated by our simulated data sets, when $\alpha^{(2)}_0 = 3$, the intercept term will determine over 95\% of the variance of $y$. This in turn reduces the efficiency with which covariates and random scale effects are modeled. In such scenarios, the errors associated with using MELS and SPLSME as the base model tend to approach the errors of using RILM as the base model. For example, in Table \ref{error_12}, if we first use SPLSME to impute $x_1$, the smallest errors reach when using MELS, SPLSME, MELS, SPLSME to impute $y$ for $\alpha^{(2)}_0 = 0, 1, 2, 3$ respectively. The corresponding improvement ratios of errors are 1.48\%, 2.43\%, 1.15\%, and 1.09\%, showing a decreasing trend. However, even when the intercept becomes dominant, as seen when $\alpha^{(2)}_0$ is set to 3, the errors associated with using MELS or SPLSME as the base model still remain slightly smaller than those obtained when using RILM as the base model. 

Apart from the change of error difference, when $\alpha^{(2)}_0$ increases, estimation of the within-variance, like the intercept $\alpha^{(2)}_0$, and the coefficient of subject-level covariate $\alpha^{(2)}_2$, will become unreliable.  For details on the exact values of this decrease, please refer to Appendix \ref{appendix3}, where the estimates for $\alpha^{(2)}_0 = 1.00$, $\alpha^{(2)}_0 = 2.00$, and $\alpha^{(2)}_0 = 3.00$ are presented. Here we highlight some trends of these estimates. For $\alpha^{(2)}_0$, its coverage by MELS decreases from 100\% to 89\%, and finally 43\%, and its coverage by SPLSME decreases from 100\% to 98\%, 81\%, and finally 31\%. For $\alpha^{(2)}_2$, its coverage of MELS decreases from 100\% to 99\%, 89\%, and finally 52\%, and its coverage of SPLSME decreases from 100\% to 95\%, 82\%, and finally 43\%. This trend can be attributed to the subject-level covariate ${x_2}$, which is designed to mimic demographic features. This covariate remains more consistent compared to ${x_1}$, which changes not only across subjects but also across occasions. Thus, the ${x_2}$ term sometimes might look like the intercept term. As the intercept of the variance model increases, MELS and SPLSME cannot distinguish the within-variance of the response caused by the intercept term and the change of ${x_2}$. Thus, they will inaccurately attribute the within-variance caused by ${x_2}$ to the intercept component or vice versa. This results in unreliable estimation of the coefficient of the intercept, $\alpha^{(2)}_0$, and $\alpha^{(2)}_2$. 

Next, we change the relationship between the random location effect and the random scale effect from a weak negative correlation into a strong negative correlation. Since RILM cannot model this correlation, we observe a slight increase in the difference ratio in Table \ref{error_12} from $\rho_{\eta_1^{(2)}, \eta_2^{(2)}} = -0.20$ to $\rho_{\eta_1^{(2)}, \eta_2^{(2)}} = -0.80$. For example, if we use MELS to impute $x_1$ firstly, the difference ratio of imputing $y$ increases from 0.83\% to 1.33\%, and if we use SPLSME to impute $x_1$ firstly, the difference ratio of imputing $y$ increases from 1.48\% to 2.05\%. Both of these examples indicate the insufficiency of using RILM as the base model when the correlation between the random effects cannot be ignored.

\begin{table*}[t]
\small{
    \centering
    \input{bias_0_1_-0.8_1_1_-0.5_0.5_0.1_0.05_0.1}}
\end{table*}

This increase in the correlation between $y$'s mean and within-variance also causes RILM's estimation of $\beta^{(2)}_1$ to have a larger bias. In Table \ref{bias_0_1_-0.8_1_1_-0.5_0.5_0.1_0.05_0.1}, as the correlation changes from -0.20 to -0.80, the coverage of RILM's estimation of $\beta^{(2)}_1$ decreases from 88\% to 81\%. When the correlation between $y$'s mean and within-variance is small, the across-occasion change in $y$'s mean primarily relies on $x_1$, the sole covariate that changes across occasions. However, as the correlation increases, the level of $y$'s within-variance becomes increasingly influential in determining the across-occasion change in $y$'s mean. As RILM can only use ${x_1}$ to model this across-occasion change, the estimation of $\beta^{(2)}_1$ becomes biased when the correlation between $y$'s mean and within-variance increases. 

In summary, initiating with RILM to impute the covariate $x_1$ results in significant errors, irrespective of the subsequent model chosen for imputing $y$. For imputing $y$, when $\alpha^{(2)}_0$ is relatively small, changes in covariates have a noticeable impact on the variance of $y_{ij}$, emphasizing the need for modeling the variance. Even when $\alpha^{(2)}_0$ is quite large and the variance remains approximately constant, using MELS or SPLSME to impute ${x_1}_{ij}$ and $y_{ij}$ also yields smaller errors compared to using RILM. In cases where the random location effect and random scale effect exhibit a high correlation, RILM's estimation of the mean model coefficients can become unreliable. Based on these findings, it is generally recommended to use MELS or SPLSME instead of RILM as the base model for multiple imputation. These models are more robust across a range of parameter values and data scenarios, making them a more reliable choice for imputing missing data.

\subsection{Modeling random missing effects}\label{MELS and SPLSME}

The difference between MELS and SPLSME is that SPLSME decomposes the random location effect and random scale effect into one missing-decided part and one missing-orthogonal part. In this subsection, we explore various combinations of the parameters $\gamma^{(2)}$ and $\delta^{(2)}$ associated with the missing-decided part. Our aim is to evaluate SPLSME's performance and see whether MELS maintains its efficacy. We assign negative values to $\gamma^{(2)}$, implying that as the level of $y$ increases, the probability of the observation being missing decreases. Conversely, for $\delta^{(2)}$, we allocate positive values. This suggests that as the fluctuation in $y$ grows, the probability of the observation being missing rises.

\begin{table*}[t]
\small{
    \centering
    \input{error_23}}
\end{table*}

The errors and their comparisons are presented in Table \ref{error_23}. Similar to Subsection \ref{RILM and MELS}, multiple imputation consistently outperforms single imputation as shown in Table \ref{error_23}. We will proceed to analyze the performance of these models specifically for multiple imputation. We also summarize the best and worst model for each method to impute $x_1$, and calculate their difference ratio.

As we increase the magnitude of how the missing random effect alters the mean of $y$, that is, as the absolute value of $\gamma^{(2)}$ grows, we observe that SPLSME progressively outperforms MELS in multiple imputations. For example, in Table \ref{error_23}, if we use SPLSME to impute the ${x_1}$ firstly, the error of MELS is essentially the same as for SPLSME, namely, 8.00 for MELS and 8.01 for SPLSME. When we increase the absolute value of $\gamma^{(2)}$, SPLSME outperforms the MELS model and the difference ratio between these two models also increases. When $\gamma^{(2)} = -1.00$, the difference ratio is $\frac{7.66 - 7.64}{7.66} \approx 0.26\%$, and when $\gamma^{(2)}$ is -1.50, the difference ratio increases to $\frac{6.67 - 6.63}{6.67} \approx 0.60\%$. Also, when the relationship between the missing effect and the mean of $y$ is strong, the performance gap between SPLSME and RILM becomes more pronounced. The error difference ratios between using RILM and SPLSME are $\frac{8.12 - 8.01}{8.12} \approx 1.35\%$, 2.68\%, 10.53\% for $\gamma = -0.50, -1.00, -1.50$ respectively. Upon analyzing the errors, we can see that when the missing effect's impact on the response's mean is large; in this case, SPLSME, by modeling the random missing effect, can exhibit an advantage in multiple imputation. 

\begin{table*}[t]
\small{
    \centering
    \input{bias_0_1_-0.2_1_1_-0.5_1_0.1_0.05_0.1}
    }
\end{table*}

As we increase the magnitude of how the missing random effects influence the variance of $y$, denoted by the parameter $\delta^{(2)}$, from 0.50 to 0.75 and 1.00, unlike for the change of $\gamma^{(2)}$, SPLSME does not exhibit a clear error decrease over MELS\@.   This discrepancy can be attributed to the fact that $\gamma^{(2)}$ governs the mean of $y$, making its impact on imputation more readily apparent compared to $\delta^{(2)}$, which primarily controls the variance of $y$. However, for MELS, some of the 
estimated parameters are unreliable.  For example, the coverage rates of $\alpha^{(2)}_1$ is 92\% when $\delta^{(2)} = 0.50$ by Table \ref{bias_0_1_-0.2_1_1_-0.5_0.5_0.1_0.05_0.1}, and decreases to 86\% when $\delta^{(2)} = 1.00$ in Table \ref{bias_0_1_-0.2_1_1_-0.5_1_0.1_0.05_0.1}. Notably, we observe an increase in the coverage of $\alpha^{(2)}_0$ in RILM. It increases from 12\% to 42\% when $\delta^{(2)}$ changes from 0.50 to 1.00. It is because RILM utilizes only $\alpha^{(2)}_0$ to model the entire expression $\alpha^{(2)}_0 + \alpha^{(2)}_1 {x_1}_{ij} + \alpha^{(2)}_2 {x_2}_i + \eta^{(2)}_{2, i} + \delta^{(2)}\lambda_i$. When the value of $\delta^{(2)}$ increases, it shifts the value of the entire expression closer to the true value of $\alpha^{(2)}_0$, which is set to 0.00 in our simulation. Consequently, the increase in coverage does not necessarily indicate an improvement in the accuracy of estimating $\alpha^{(2)}_0$. 

Upon analyzing the impact of $\delta^{(2)}$, we can see that while SPLSME may not lead to a substantial reduction in errors as $\delta^{(2)}$ increases, it does provide estimators that are less biased and more consistent.

When examining the estimate of the random location effect ($\sigma^{(2)}_{v_1}$) and the correlation between the random location and scale effect ($\rho_{v_1^{(2)}, v_2^{(2)}}$), we find that the bias and coverage for the RILM and MELS models are acceptable for all $\gamma^{(2)}$ and $\delta^{(2)}$ values. This suggests that while RILM and MELS cannot differentiate between the variances of random effects attributable to missing-decided or missing-orthogonal components, they still provide good estimation of the overall variance and correlation of random effects. In Appendix \ref{appendix3}, we provide the estimate regarding $\gamma^{(2)} = -1.00$, $\gamma^{(2)} = -1.50$, and $\delta^{(2)} = 0.75$. They also show that RILM and MELS provide good estimation of the overall variance and correlation of random effects.

In summary, SPLSME exhibits a slight advantage over MELS in terms of reducing errors in multiple imputation or improving the accuracy of estimators when the impact of the missing data on $y$ is large. When the 
effect of the missing data on $y$ is relatively minor, the performance of MELS is comparable to that of SPLSME.

\section{Multiple imputation on caregivers' emotion study}\label{example}
\subsection{Models}
In this section, we will conduct multiple imputation on the caregivers data set introduced in Section \ref{motivating example}. In the data set, two variables have missing observations: positive affect (PA) and stress. We will treat stress as the response and illustrate how to impute stress by the proposed multiple imputation method.

In the model of mean and variance, all available covariates will be included: age, level of education, dementia duration in years, Clinical Dementia Rating (CDR), and PA. For the level of education, two dummy variables
will be used, denoted as ${\text{edu}_2}_i$ and ${\text{edu}_3}_i$, to signify education levels 2 and 3, respectively. Similarly, for CDR, two dummy variables, ${\text{CDR}_2}_i$ and ${\text{CDR}_3}_i$, represent CDR levels 2 and 3. In the model of missing, the sequential day and beep within data will be used as covariates. 
The sequential day number will be treated as continuous a variable. Regarding the beep number, four dummy variables will be used , ${\text{beep}_{3, 4}}_{ij}$, ${\text{beep}_{5, 6}}_{ij}$, ${\text{beep}_{7, 8}}_{ij}$, and ${\text{beep}_{9, 10}}_{ij}$, to indicate whether the beep number falls within the ranges of 3-4, 5-6, 7-8, and 9-10, respectively.

To impute stress, it is necessary to impute PA first. The covariates we employ to impute PA are largely the same as those for stress. However, since in this data set, the demographic features do not have much correlation with PA, and thus have little effect on the imputation of PA, we introduce the day number as one of the covariates in the model for both the mean and variance of PA. This adjustment ensures a more effective imputation process for PA and, consequently, stress.

The same three models as introduced before — RILM, MELS, and SPLSME— will be used to sequentially impute missing values for PA and stress. 

\subsection{ Parameter estimates}

Before doing the imputation, we firstly fit the three models for PA as response and stress as response. The parameter estimates for PA are presented in Table \ref{example_PA}, and the estimates for stress are presented in Table \ref{example_stress}.

\begin{table}[t]
\small{
    \centering
    \input{example_PA}
    }
\end{table}

\begin{table}[t]
\small{
    \centering
    \input{example_stress}
    }
\end{table}

In the mean model for PA, by observing the intervals, we can see that most of the results do not reach statistical significance. An exception is $\beta_{day}$, which exhibits a significant negative effect in the RILM model but does not quite reach significance in the MELS or SPLSME models. From the simulation
study discussed in Section \ref{simulation study}, when there is a substantial correlation between the random location and scale effects, the estimates in the RILM model of covariates that change at both BS and WS levels 
(i.e., time varying covariates) might become unreliable. In the present case, the correlation between random location and scale effects is estimated as -0.43 in the MELS model, and -0.44 in SPLSME, and is significant in both models. Consequently, this might explain some of the difference observed for $\beta_{day}$ between the models.

In the model for the variance of PA, $\alpha_{day}$ achieves statistical significance at the 5\% alpha level in both the MELS and SPLSME models. Its estimated values are -0.12 for both models, suggesting that caregivers tend to experience less variability in PA as the experiment day increases. RILM's estimation of the intercept, $\alpha_{inter}$, significantly diverges from that of both MELS and SPLSME. This discrepancy arises because RILM only includes this parameter to model variance, whereas it represents the variance when covariates equal 0 
in the other two models.  

In the model of missingness, the estimates of $\tau_{beep_{3, 4}}$, $\tau_{beep_{5, 6}}$, $\tau_{beep_{7, 8}}$, and $\tau_{beep_{9, 10}}$ are all negative and achieve statistical significance. This suggests that missing data tends to occur more frequently during the early morning beeps, and less so as the day goes on. 
Furthermore, the estimate of $\gamma$ is -0.96, and is observed to be statistically significant. 
This implies that caregivers with higher PA values experience fewer missing data points. 
In other words, individuals with better moods are perhaps more willing to respond to the beeps. 
Missingness accounts for  $\frac{(-0.96 * 0.57)^2}{(-0.96 * 0.57)^2 + (0.78)^2} \approx 32.98\%$ of the variance of the random location effect and $\frac{(0.34 * 0.57)^2}{(0.34 * 0.57)^2 + (0.70)^2} \approx 7.12\%$ of the variance of the random scale effect. In Section \ref{simulation study}, it was shown that if missing data significantly impacts the response's mean, the SPLSME method outperforms MELS in slightly reducing imputation errors. This indicates that employing SPLSME might be advantageous than MELS when conducting multiple imputations for PA.

The Bayesian Information Criterion (BIC) \cite{neath2012bayesian} was computed for the three models, and are listed towards the bottom of Table \ref{example_PA}. The BIC suggests that the model's fit significantly improves when transitioning from RILM to MELS. However, introducing additional parameters from MELS to SPLSME might not be necessary. It is worth noting that the BIC relies on the assumption that the model is fitted using Maximum Likelihood Estimation (MLE), but the estimation for MELS and SPLSME do not adhere to MLE. Thus, we further evaluated the models using the Expected Log Predictive Density (ELPD) as a measure of predictive performance \cite{vehtari2017practical}. ELPD is calculated through the leave-one-out cross-validation, and a higher ELPD indicates better predictive performance. In our Table \ref{example_PA}, the EPLS values for both MELS and SPLSME are comparable, suggesting that these two methods exhibit similar performance.

Estimates for stress are presented in Table \ref{example_stress}. 
In all casess, $\beta_{PA}$ and $\alpha_{PA}$ are negative and observed to be statistically significant.
This suggests that caregivers with higher PA tend to experience lower stress levels and reduced stress fluctuations. For the MELS and SPLSME models, $\alpha_{age}$ is estimated to be -0.04 and statistically significant. This suggests that older caregivers tend to have lower within-variance of stress. 

The estimates for the missing-related parameters are close to those in the PA model. This similarity arises because PA and stress data are missing concurrently, indicating identical missing patterns. Their minor differences are likely due to the inherent randomness in the MCMC approach used for estimation. 
In the SPLSME model, the estimate for $\delta$ is -0.44 and statistically significant. This implies that caregivers with greater within-subject variance in stress are less likely to have missing data. 
In terms of variance explained, missingness accounts for $\frac{(-0.04 * 0.57)^2}{(-0.04 * 0.57)^2 + (0.54)^2} \approx 0.18\%$ of the variance of the random location effect and about $\frac{(-0.44 * 0.57)^2}{(-0.44 * 0.57)^2 + (0.36)^2} \approx 32.68\%$ of the variance of random scale effect. Thus, missingness is minimally related to 
the random location effect, but does appear to be moderately related to the random scale effect.  
As in the analysis of PA, the ELPD of the MELS and SPLSME are essentially equal. This suggests that 
the additional modeling of missingness that the SPLSME model includes might not be necessary in this particular case. 

\subsection{Visualizations}

After fitting each three models for PA and stress, we then conduct the multiple imputation. The imputation results will be shown and analyzed in this subsection.

\begin{figure*}[ht]
  \centering
  \includegraphics[width=\textwidth]{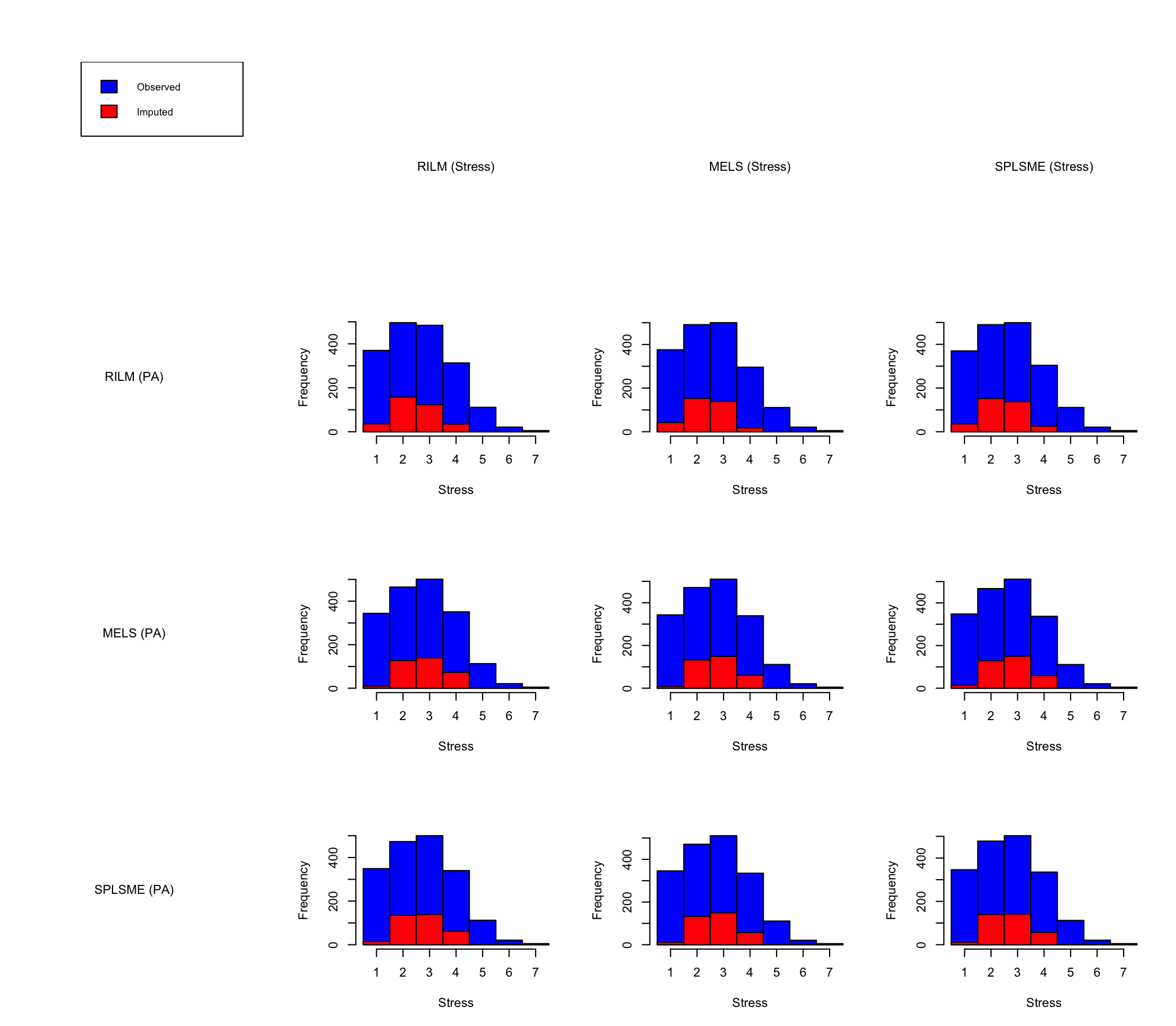}
  \caption{Histograms depicting observed and imputed stress values across three models: Rows 1, 2, and 3 correspond to the use of RILM, MELS, and SPLSME for imputing PA, respectively; Columns 1, 2, and 3 represent the use of RILM, MELS, and SPLSME for imputing stress, respectively.}
  \label{hist}
\end{figure*}

To represent the imputed continuous stress values as ordinal values ranging from 1 to 7, cutoffs at 1.5, 2.5, 3.5, 4.5, 5.5, and 6.5 were used. Figure \ref{hist} displays histograms illustrating the observed and imputed values of stress. In these histograms, the red segments represent the frequency of the imputed values, while the blue segments represent the frequency of the observed values. Rows 1, 2, and 3 correspond to the utilization of the RILM, MELS, and SPLSME methods, respectively, for imputing the PA, while columns 1, 2, and 3 pertain to the application of these methods for imputing stress. Notably, our findings reveal a common trend across all models, with a concentration of values around 2 and 3, and a relative scarcity of values on both ends of the distribution. However, there are some slight differences in the imputation results produced by different models. For instance, RILM imputation tends to assign more values to 2 than 3, while MELS and SPLSME exhibit the opposite pattern, though the differences are relatively minor.

\begin{figure*}[ht]
  \centering
  \includegraphics[width=\textwidth]{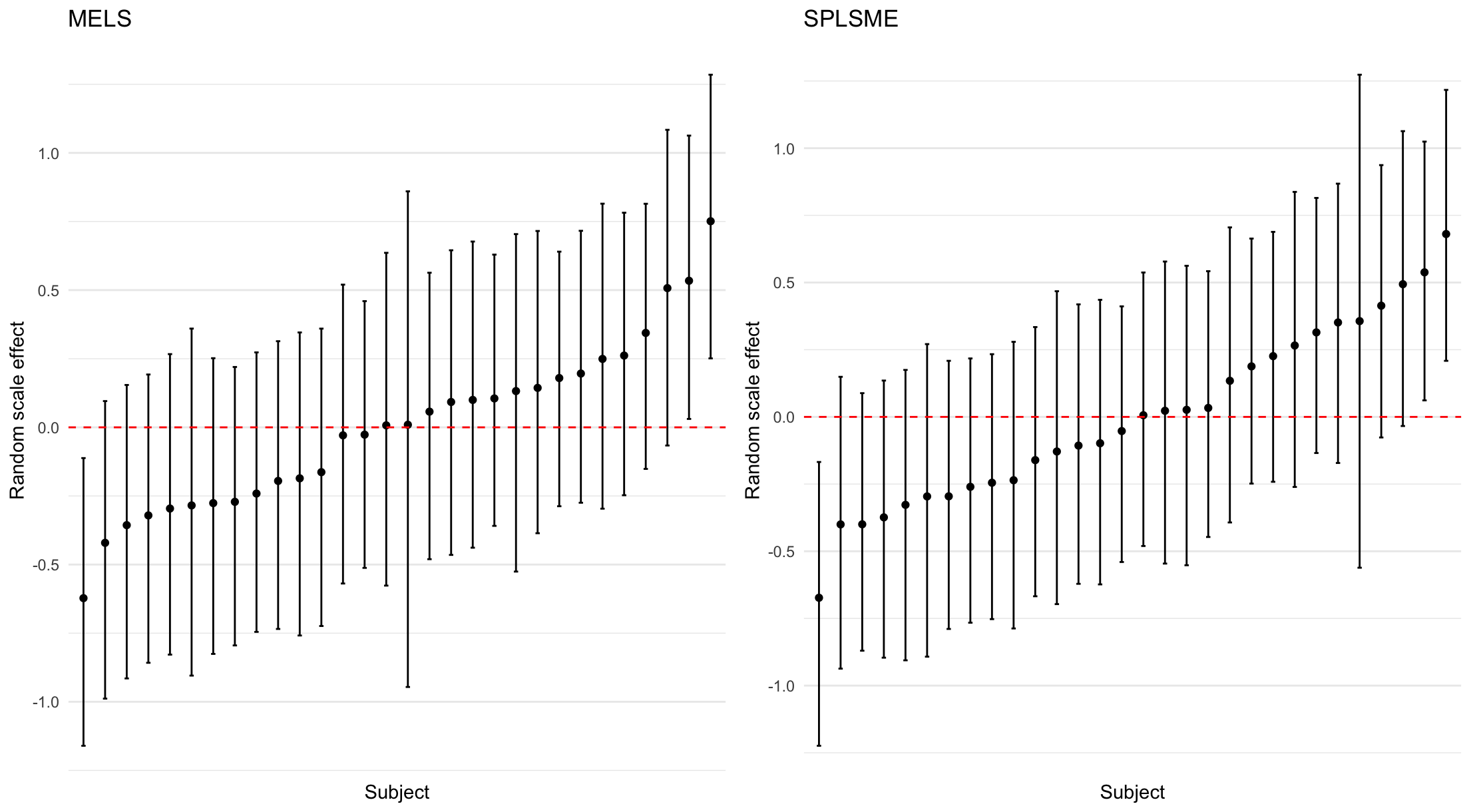}
  \caption{Estimated random scale effects of different subjects by MELS and SPLSME}
  \label{caterpillar}
\end{figure*}

We generated caterpillar plots in Figure \ref{caterpillar}, to illustrate the random scale effects in the model of stress for all subjects. Since the RILM doesn't model the random scale effects, Figure \ref{caterpillar} only shows the random scale effects estimated by MELS and SPLSME. On the plot, points to the right of the center line represent subjects with random effects larger than the average, while points to the left indicate smaller random scale effects. The varying lengths of the confidence intervals around these points signify differences in the variability of the random scale effects. The caterpillar plot demonstrates the heterogeneity across subjects in their degree of WS variance, emphasizing the inaccuracy of homogeneous WS variance for all subjects, as in the RILM model.

\begin{figure*}[ht]
  \centering
  \includegraphics[width=\textwidth]{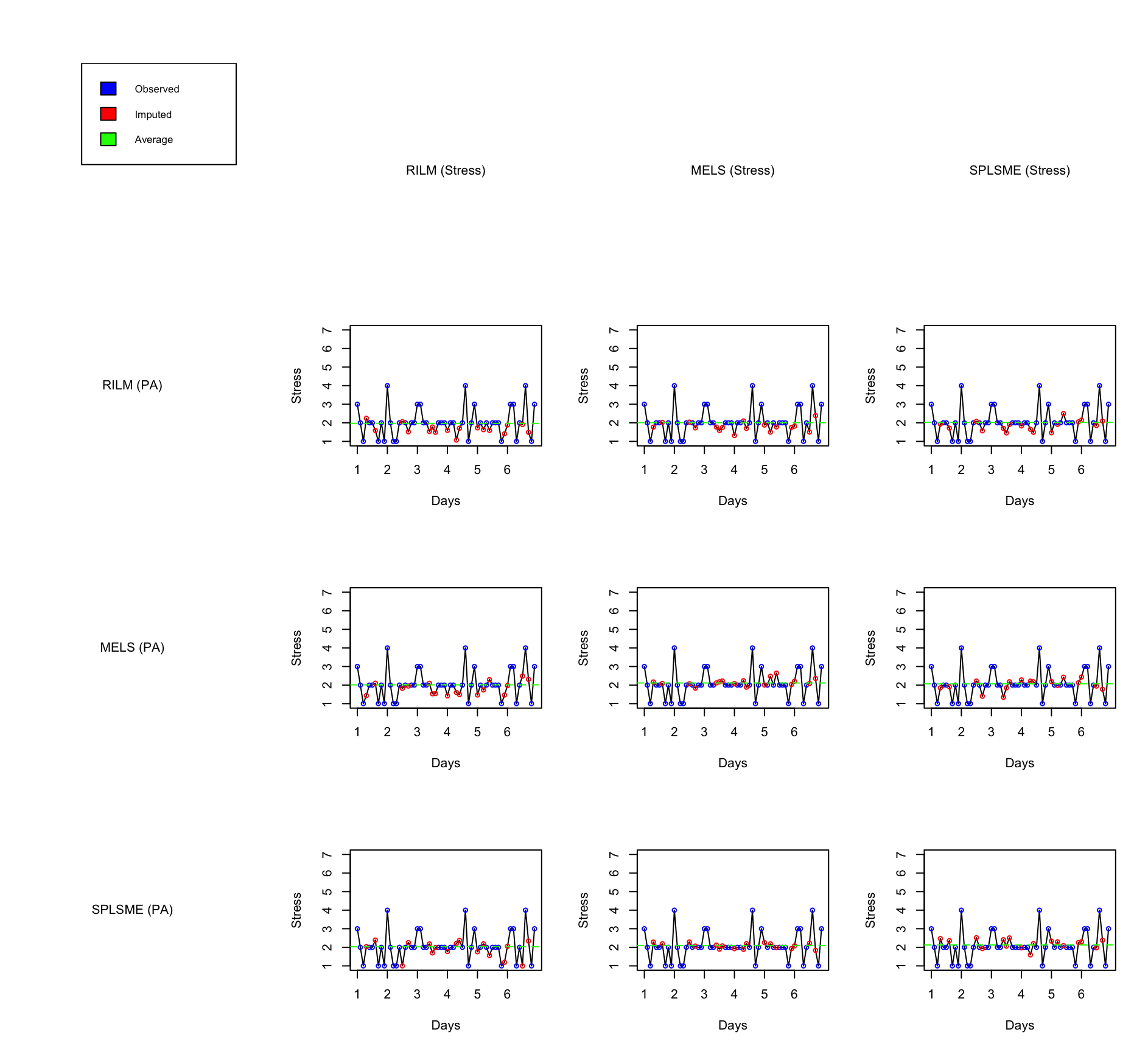}
  \caption{Line charts depicting observed and imputed stress values across three models for the 8th subject: Rows 1, 2, and 3 correspond to the use of RILM, MELS, and SPLSME for imputing PA, respectively; Columns 1, 2, and 3 represent the use of RILM, MELS, and SPLSME for imputing stress, respectively.}
  \label{8subj}
\end{figure*}

The leftmost points on both caterpillar plots (i.e., point with the lowest value) correspond to the random scale effect estimates of the 8th subject. This indicates that the unexplained WS variance of subject 8 is the lowest among all the subjects. In Figure \ref{8subj}, we present the multiple imputation results for this subject using different models. Rows 1, 2, and 3 represent the application of the RILM, MELS, and SPLSME methods, respectively, for imputing PA. Columns 1, 2, and 3 represent the use of these methods for imputing stress levels. In the figure, red points denote imputed values, while blue points denote observed values. Additionally, a green line is included to display the average stress values. When considering the modeling of the random scale effect, the imputation results in columns 2 and 3, generated by MELS and SPLSME, demonstrate greater consistency around the green line. For example, there are two missing observations between the end of the 5th day and the beginning of the 6th day. The two observed values flanking these missing observations are 1 and 3, situated to the left and right, respectively. All models produce imputed values between 1 and 3. However, the first imputed value in the first column is closer to 1, which significantly deviates from the average line, approximately around 2. In contrast, for the second and third columns, the imputed first values cluster around 2, displaying better alignment with the average line.

In summary, the analysis presented above serves as an illustrative example of the practical application of our multiple imputation method to a real data set. It demonstrates the importance of selecting the most appropriate posterior model based on the data set's specific characteristics. The choice of different posterior models yields distinct multiple imputation results, and two key considerations should guide this choice: the impact of covariates as well as random effects on the within-variance, and the magnitude of missing nature in relation to the response variable. By carefully considering these factors, researchers can make informed decisions when selecting a posterior model for multiple imputations, ultimately leading to a more accurate and robust imputed data set for further analysis and inference.

\section{Discussion}\label{discussion}
In this paper, we adopt a two-step Bayesian approach for multiple imputation: by assuming a posterior distribution for the data $y$ given a parameter set $\theta$, we firstly derive the posterior distribution of $\theta$ based on the observed data and draw $m$ different $\theta$ values from this posterior distribution, then we draw $m$ imputed values from the posterior distribution of $y$, conditioning on each of the $\theta$ values. We have employed three distinct models for longitudinal data, namely the Random Intercept Linear Mixed (RILM) model, the Mixed-effects Location Scale (MELS) model, and the Shared Parameter Location Scale Mixed Effect (SPLSME) model to complete the posterior distribution within this framework. The MELS model extends the RILM by accommodating changes in the within-subject variance based on covariates and random effects, and the SPLSME model further builds upon the MELS framework by incorporating a random intercept logistic model for intermittent missing data, establishing a link between missing data and the response variable. For model estimation in both MELS and SPLSME, we employ a Markov Chain Monte Carlo (MCMC) approach to alleviate the computational challenges associated with multidimensional numerical integration in Maximum Likelihood Estimation (MLE). Given the characteristics of the EMA data set, which frequently exhibits simultaneously missing variables due to participant nonresponse to scheduled prompts, our multiple imputation methodology is designed to proceed in two steps: we first impute the missing covariates and then impute the response variable using both observed data and the imputed covariates.

Simulation results under the Missing Not At Random (MNAR) assumption reveal a clear advantage of multiple imputations over single imputations. Of the models employed for multiple imputations, both MELS and SPLSME outperform the RILM, particularly when covariates and random effects exert a substantial influence on the within-variance of the response variable. It is worth noting that the performance of MELS and SPLSME, while generally similar, may diverge when the extent of missing data is largely related to observations. Within the SPLSME model, two parameters, $\gamma$ and $\delta$, are set to quantify the relationship between missing data and the subject's mean and within-variance, respectively. A large absolute value of $\gamma$ indicates a significant association between missing data and the subject's mean, leading to smaller multiple imputation errors in SPLSME compared to MELS. Similarly, a large absolute value of $\delta$ signifies a strong connection between missing data and the subject's within-variance, resulting in more accurate model estimations within the SPLSME framework compared to MELS.

Subsequently, we applied our multiple imputation techniques to the caregivers' emotion study and provided visual representations of the results. We also conducted an analysis of the model selection process for multiple imputations. We propose that two key considerations should guide this selection: the influence of covariates as well as random effects on the within-variance, and the nature of missing data in relation to the response variable.

While our exploration has predominantly concentrated on the conventional and widely studied normal distribution, within the same Bayesian framework, our multiple imputation method can be readily extended to accommodate other types of outcomes, including binary and Poisson outcomes. These extensions can significantly broaden the applicability of our method to a wider range of research scenarios, where the nature of the data may not conform to a normal distribution. 

Due to the nature of the EMA experiment, many EMA data sets exhibit simultaneous missingness across multiple variables. In this study, we employ a step-by-step technique to impute these multiple missing variables sequentially. However, such a step-by-step training and imputation process can be time-consuming. It would be beneficial to obtain the joint distribution of all the missing variables and impute them simultaneously. One possible method is to use the Gibbs sampling technique, which iteratively samples from the full conditional distributions for each variable. Future work could entail further exploration in this area.

%Bibliography
\bibliographystyle{unsrt}  
\bibliography{references} 

\appendix

\section{First Appendix: Multiple imputation sampling for three models}\label{appendix1}
In Section \ref{statistical analysis}, we present the general two-step Bayesian approach used for our multiple imputation. In this appendix, we delve into the specifics of each step, tailored to the different settings of the three models.

\subsection{RILM}

In the first step, sample $m$ parameter sets $\{\hat\beta_0^{(k)}, \hat\beta^{(k)}, \hat\alpha_0^{(k)}, \{\hat v^{(k)}_{1, i}\}_{i = 1, \dots, n} \}_{k = 1, \dots, m}$ from the posterior distribution of parameters conditional on observed data, and calculate

\begin{align}
&\hat\mu^{(k)}_{ij} = \hat\beta_0^{(k)} + x_{ij}^\top \hat\beta^{(k)} + \hat v^{(k)}_{1, i}  \\
&{{\hat{\sigma^2}}_{ij}}^{(k)} = e^{ \hat\alpha_0^{(k)}}.
\end{align}

In the second step, we sample

\begin{align}
\hat y^{(k)}_{ij} \sim \mathcal N 
\left( \hat \mu^{(k)}_{ij}, {\hat{\sigma^2}_{ij}}^{(k)} \right) .
\end{align}

By repeating sampling $m$ times for each parameter set, we get $m$ imputation values $\{\hat y^{(k)}_{ij}\}_{k = 1, \dots, m}$.

\subsection{MELS}

In the first step, sample $m$ parameter sets $\{\hat\beta_0^{(k)}, \hat\beta^{(k)}, \hat\alpha_0^{(k)}, \hat\alpha^{(k)}, \{\hat v^{(k)}_{1, i}\}_{i = 1, \dots, n}, \{\hat v^{(k)}_{2, i}\}_{i = 1, \dots, n}\}_{k = 1, \dots, m}$ from the posterior distribution of parameters conditional on observed data, and calculate

\begin{align}
&\hat\mu^{(k)}_{ij} = \hat\beta_0^{(k)} + x_{ij}^\top \hat\beta^{(k)} + \hat v^{(k)}_{1, i}  \\
&{{\hat{\sigma^2}}_{ij}}^{(k)} = e^{ \hat\alpha_0^{(k)} + x_{ij}^\top \hat\alpha^{(k)} + \hat v^{(k)}_{2, i} }.
\end{align}

In the second step, sample

\begin{align}
\hat y^{(k)}_{ij} \sim \mathcal N 
\left( \hat \mu^{(k)}_{ij}, {\hat{\sigma^2}_{ij}}^{(k)} \right) .
\end{align}

By repeating sampling $m$ times for each parameter set, we get $m$ imputation values $\{\hat y^{(k)}_{ij}\}_{k = 1, \dots, m}$.

\subsection{SPLSME}

In the first step, sample $m$ parameter sets $\{\hat\beta_0^{(k)}, \hat\beta^{(k)}, \hat\alpha_0^{(k)}, \hat\alpha^{(k)}, \hat\gamma^{(k)}, \hat\delta^{(k)}, \{\hat\eta^{(k)}_{1, i}\}_{i = 1, \dots, n}, \{\hat\eta^{(k)}_{2, i}\}_{i = 1, \dots, n}, \{\hat\lambda^{(k)}_i \}_{i = 1, \dots, n} \}_{k = 1, \dots, m}$ from the posterior distribution of parameters conditional on observed data, and calculate

\begin{align}
&\hat\mu^{(k)}_{ij} = \hat\beta_0^{(k)} + x_{ij}^\top \hat\beta^{(k)} + \hat\eta^{(k)}_{1, i} + \hat\gamma^{(k)} \hat\lambda^{(k)}_i \\
&{{\hat{\sigma^2}}_{ij}}^{(k)} = e^{\hat\alpha_0^{(k)} + x_{ij}^\top \hat\alpha^{(k)} + \hat\eta^{(k)}_{2, i} + \hat\delta^{(k)} \hat\lambda^{(k)}_i}.
\end{align}

In the second step, sample

\begin{align}
\hat y^{(k)}_{ij} \sim \mathcal N 
\left( \hat \mu^{(k)}_{ij}, {\hat{\sigma^2}_{ij}}^{(k)} \right) .
\end{align}

By repeating sampling $m$ times for each parameter set, we get $m$ imputation values $\{\hat y^{(k)}_{ij}\}_{k = 1, \dots, m}$.

\section{Second Appendix: Bayesian estimation for MELS and SPLSME}\label{appendix2}
For estimating MELS and SPLSME, we employ the MCMC method to converge to the posterior distribution of the parameter set $\theta$. In Section \ref{statistical analysis}, we outline a general methodology for this estimation. This appendix further elaborates on the details of the two models for the Algorithm \ref{alg1}.

\subsection{MELS}
Remind that the model setting of MELS is:

\begin{align}
&y_{ij} \mid x_{ij}, v_{1, i}, v_{2, i} \sim \mathcal{N}
\left(
\beta_0 + x^\top_{ij} \beta + v_{1, i} , e^{\alpha_0 + x_{ij}^\top \alpha + v_{2, i}}
\right)\\
&\begin{pmatrix}
    v_{1, i} \\ v_{2, i}
\end{pmatrix}
\sim
\mathcal{N}
\left(
\begin{pmatrix}
    0 \\ 0
\end{pmatrix}
,
\begin{pmatrix}
    \sigma_{v_{1}}^2 & \rho_{v_{1}, v_{2}}\sigma_{v_{1}}\sigma_{v_{2}}\\
    \rho_{v_{1}, v_{2}}\sigma_{v_{1}}\sigma_{v_{2}} & \sigma_{v_{2}}^2\\
\end{pmatrix}
\right)
.
\end{align}

Firstly we adopt the Cholesky factorization to decompose the covariance matrix and estimate the variance and correlation of random effects indirectly. By Cholesky factorization, we can decompose the covariance matrix, which is a positive-definite matrix, as the multiplication of a lower-diagonal matrix and its transpose:

\begin{align}
\begin{pmatrix}
    v_{1, i} \\ v_{2, i} 
\end{pmatrix}
=
\begin{pmatrix}
    s_{11} & 0 \\
    s_{21} & s_{22} 
\end{pmatrix}
\begin{pmatrix}
    z_{1, i} \\ z_{2, i}
\end{pmatrix},
\end{align}
where
\begin{align}
&s_{11} = \sigma_{v_{1}} \\
&s_{21} = \sigma_{v_{2}} \rho_{v_{1}, v_{2}} \\
&s_{22} = \sqrt{\sigma^2_{v_{2}} - \left(\sigma_{v_{2}} \rho_{v_{1}, v_{2}}\right)^2}  \\
&\begin{pmatrix}
z_{1, i} \\ z_{2, i} 
\end{pmatrix}
\sim
\mathcal{N}
\left(0_2, I_2\right)
.
\end{align}

After the decomposition, the sets of parameters we want to estimate is

\begin{align}
&\theta = \left(\beta_0, \beta, \alpha_0, \alpha, s_{11}, s_{21}, s_{22}, \{z_{1, i}\}_{i = 1, \dots, n}, \{z_{2, i}\}_{i = 1, \dots, n} \right)  .
\end{align}

In MELS model,$y_{obs}$ is the non-missing response $y_{ij}$. Thus, the likelihood $L(y_{obs} \mid \theta)$ can be explicitly formulated based on the normal distribution model configuration of MELS.

\subsection{SPLSME}

Remind that the model setting of SPLSME is:

\begin{align}
&y_{ij} \mid x_{ij}, \eta_{1, i}, \eta_{2, i}, \lambda_i  \sim \mathcal{N}
\left(
\beta_0 + x^\top_{ij} \beta + \eta_{1, i} + \gamma \lambda_i ,
e^{\alpha_0 + x_{ij}^\top \alpha + \eta_{2, i} + \delta \lambda_i}
\right)
\\
&m_{ij} \mid t_{ij}, \lambda_i \sim \mathcal{B}
\left(
L\left(\tau_0 + t_{ij}^\top \tau + \lambda_i
\right)\right)
\\
&\begin{pmatrix}
    \eta_{1, i} \\ \eta_{2, i} \\ \lambda_i
\end{pmatrix}
\sim
\mathcal{N}
\left(
\begin{pmatrix}
    0 \\ 0 \\ 0
\end{pmatrix}
,
\begin{pmatrix}
    \sigma_{\eta_{1}}^2 & \rho_{\eta_{1}, \eta_{2}}\sigma_{\eta_{1}}\sigma_{\eta_{2}} & 0\\
    \rho_{\eta_{1}, \eta_{2}}\sigma_{\eta_{1}}\sigma_{\eta_{2}} & \sigma_{\eta_{2}}^2 & 0\\
    0 & 0 & \sigma^2_{\lambda}
\end{pmatrix}
\right) 
.
\end{align}

Firstly we adopt the Cholesky factorization to decompose the covariance matrix and estimate the variance and correlation of random effects indirectly. By Cholesky factorization, we can decompose the covariance matrix, which is a positive-definite matrix, as the multiplication of a lower-diagonal matrix and its transpose:

\begin{align}
\begin{pmatrix}
    \eta_{1, i} \\ \eta_{2, i} \\ \lambda_i
\end{pmatrix}
=
\begin{pmatrix}
    s_{11} & 0 & 0 \\
    s_{21} & s_{22} & 0 \\
    0 & 0 & s_{33}
\end{pmatrix}
\begin{pmatrix}
    z_{1, i} \\ z_{2, i} \\ z_{3, i}
\end{pmatrix},
\end{align}
where
\begin{align}
&s_{11} = \sigma_{\eta_{1}} \\
&s_{21} = \sigma_{\eta_{2}} \rho_{\eta_{1}, \eta_{2}} \\
&s_{22} = \sqrt{\sigma^2_{\eta_{2}} - \left(\sigma_{\eta_{2}} \rho_{\eta_{1}, \eta_{2}}\right)^2} \\
&s_{33} = \sigma_{\lambda} \\
&\begin{pmatrix}
z_{1, i} \\ z_{2, i} \\ z_{3, i}
\end{pmatrix}
\sim
\mathcal{N}
\left(0_3, I_3\right)
.
\end{align}

After the decomposition, the set of parameters we want to estimate is

\begin{align}
&\theta_1 = \left(\beta_0, \beta, \alpha_0, \alpha, \tau_0, \tau, \gamma, \delta , s_{11}, s_{21}, s_{22}, s_{33}, \{z_{1, i}\}_{i = 1, \dots, n}, \{z_{2, i}\}_{i = 1, \dots, n}, \{z_{3, i}\}_{i = 1, \dots, n} \right) .
\end{align}

Denote $m_{ij}$ as the indicator of whether the observation is missing. In the MELS model,$y_{obs}$ contains both the non-missing response $y_{ij}$ and indicator $m_{ij}$. Since the outcomes of $y_{ij}$ and $m_{ij}$ are independent, we can write the likelihood for subject $i$, occasion $j$ as $L\left(y_{ij} \mid \theta\right) L\left(m_{ij} \mid \theta\right)$. Each $L\left(y_{ij} \mid \theta\right)$ and $ L\left(m_{ij} \mid \theta\right)$ can be explicitly formulated based on the normal distribution and Bernoulli distribution model configuration of SPLSME.

\section{Third Appendix: Tables}\label{appendix3}
This appendix contains tables detailing estimates, biases, and coverage for parameter combinations that are not included in the main text.
\begin{table*}[h]
\small{
    \centering
    \input{bias_1_1_-0.2_1_1_-0.5_0.5_0.1_0.05_0.1}
    }
\end{table*}

\begin{table*}[ht]
\small{
    \centering
    \input{bias_2_1_-0.2_1_1_-0.5_0.5_0.1_0.05_0.1}
    }
\end{table*}

\begin{table*}[ht]
\small{
    \centering
    \input{bias_3_1_-0.2_1_1_-0.5_0.5_0.1_0.05_0.1}
    }
\end{table*}

\begin{table*}[ht]
\small{
    \centering
    \input{bias_0_1_-0.2_1_1_-1_0.5_0.1_0.05_0.1}
    }
\end{table*}

\begin{table*}[ht]
\small{
    \centering
    \input{bias_0_1_-0.2_1_1_-1.5_0.5_0.1_0.05_0.1}
    }
\end{table*}

\begin{table*}[ht]
\small{
    \centering
    \input{bias_0_1_-0.2_1_1_-0.5_0.75_0.1_0.05_0.1}
    }
\end{table*}

\end{document}

%% file: error_12.tex
\caption{Imputation errors with different $\alpha^{(2)}_0$ (the intercept of the within-variance) and $\rho_{\eta_{1}^{(2)}, \eta_{2}^{(2)}}$ (the correlation between residual random location and random scale effects)}
\label{error_12}
\begin{threeparttable}
\begin{tabular}{lllllllllll}
\hline
\multicolumn{2}{c}{\textbf{Method to impute $x_1$}}  & \multicolumn{9}{c}{\textbf{RILM}}                                                                            \\ \hline
\multicolumn{2}{c}{\textbf{Method to impute $y$}}    & \multicolumn{2}{c}{\textbf{RILM}}                                           & \multicolumn{2}{c}{\textbf{MELS}}                                           & \multicolumn{2}{c}{\textbf{SPLSME}}                                         & \multicolumn{1}{c}{\multirow{2}{*}{\textbf{Best}}}\tnote{a} & \multicolumn{1}{c}{\multirow{2}{*}{\textbf{Worst}}}\tnote{b} & \multicolumn{1}{c}{\multirow{2}{*}{\textbf{Difference  (\%)}}}\tnote{c} \\ \cline{1-8}
\multicolumn{2}{c}{\textbf{Imputation method}}       & \multicolumn{1}{c}{\textbf{Single}} & \multicolumn{1}{c}{\textbf{Multiple}} & \multicolumn{1}{c}{\textbf{Single}} & \multicolumn{1}{c}{\textbf{Multiple}} & \multicolumn{1}{c}{\textbf{Single}} & \multicolumn{1}{c}{\textbf{Multiple}} & \multicolumn{1}{c}{}                               & \multicolumn{1}{c}{}                                & \multicolumn{1}{c}{}                                           \\ \hline
$\alpha^{(2)}_0$ & $\rho_{\eta_{1}^{(2)}, \eta_{2}^{(2)}}$ &                                     &                                       &                                     &                                       &                                     &                                       &                                                    &                                                     &                                                                \\ \hline
0.00             & -0.20                             & 11.11                               & 9.51                                  & 9.98                                & 9.35                                  & 10.10                               & 9.40                                  & MELS                                               & RILM                                                & 1.68                                                           \\ \hline
1.00             & -0.20                             & 15.54                               & 11.51                                 & 13.51                               & 11.80                                 & 13.58                               & 11.79                                 & RILM                                               & RILM                                                & 2.46                                                           \\ \hline
2.00             & -0.20                             & 20.20                               & 15.31                                 & 19.77                               & 15.24                                 & 20.13                               & 15.31                                 & RILM                                               & MELS                                                & 0.46                                                           \\ \hline
3.00             & -0.20                             & 38.10                               & 25.82                                 & 39.42                               & 25.83                                 & 39.76                               & 25.83                                 & MELS                                               & RILM                                                & 0.04                                                           \\ \hline
0.00             & -0.80                             & 12.47                               & 10.30                                 & 10.92                               & 10.21                                 & 11.00                               & 10.25                                 & MELS                                               & RILM                                                & 0.87                                                           \\ \hline
\multicolumn{2}{c}{\textbf{Method to impute $x_1$}}  & \multicolumn{9}{c}{\textbf{MELS}}                                                                                                                                                                                                                                                                                                                                                                                   \\ \hline
\multicolumn{2}{c}{\textbf{Method to impute $y$}}    & \multicolumn{2}{c}{\textbf{RILM}}                                           & \multicolumn{2}{c}{\textbf{MELS}}                                           & \multicolumn{2}{c}{\textbf{SPLSME}}                                         & \multicolumn{1}{c}{\multirow{2}{*}{\textbf{Best}}} & \multicolumn{1}{c}{\multirow{2}{*}{\textbf{Worst}}} & \multicolumn{1}{c}{\multirow{2}{*}{\textbf{Difference  (\%)}}} \\ \cline{1-8}
\multicolumn{2}{c}{\textbf{Imputation method}}       & \multicolumn{1}{c}{\textbf{Single}} & \multicolumn{1}{c}{\textbf{Multiple}} & \multicolumn{1}{c}{\textbf{Single}} & \multicolumn{1}{c}{\textbf{Multiple}} & \multicolumn{1}{c}{\textbf{Single}} & \multicolumn{1}{c}{\textbf{Multiple}} & \multicolumn{1}{c}{}                               & \multicolumn{1}{c}{}                                & \multicolumn{1}{c}{}                                           \\ \hline
$\alpha^{(2)}_0$ & $\rho_{\eta_{1}^{(2)}, \eta_{2}^{(2)}}$&                                     &                                       &                                     &                                       &                                     &                                       &                                                    &                                                     &                                                                \\ \hline
0.00             & -0.20                             & 10.37                               & 8.40                                  & 8.94                                & 8.33                                  & 9.00                                & 8.36                                  & MELS                                               & RILM                                                & 0.83                                                           \\ \hline
1.00             & -0.20                             & 14.03                               & 10.50                                 & 12.23                               & 10.41                                 & 12.18                               & 10.38                                 & SPLSME                                             & RILM                                                & 1.14                                                           \\ \hline
2.00             & -0.20                             & 19.10                               & 14.28                                 & 18.67                               & 14.21                                 & 19.14                               & 14.22                                 & MELS                                               & RILM                                                & 0.49                                                           \\ \hline
3.00             & -0.20                             & 36.90                               & 24.85                                 & 37.90                               & 24.79                                 & 37.76                               & 24.75                                 & SPLSME                                             & RILM                                                & 0.40                                                           \\ \hline
0.00             & -0.80                             & 11.21                               & 9.03                                  & 9.59                                & 8.91                                  & 9.61                                & 8.95                                  & MELS                                               & RILM                                                & 1.33                                                           \\ \hline
\multicolumn{2}{c}{\textbf{Method to impute $x_1$}}  & \multicolumn{9}{c}{\textbf{SPLSME}}                                                                          \\ \hline
\multicolumn{2}{c}{\textbf{Method to impute $y$}}    & \multicolumn{2}{c}{\textbf{RILM}}                                           & \multicolumn{2}{c}{\textbf{MELS}}                                           & \multicolumn{2}{c}{\textbf{SPLSME}}                                         & \multicolumn{1}{c}{\multirow{2}{*}{\textbf{Best}}} & \multicolumn{1}{c}{\multirow{2}{*}{\textbf{Worst}}} & \multicolumn{1}{c}{\multirow{2}{*}{\textbf{Difference  (\%)}}} \\ \cline{1-8}
\multicolumn{2}{c}{\textbf{Imputation method}}       & \multicolumn{1}{c}{\textbf{Single}} & \multicolumn{1}{c}{\textbf{Multiple}} & \multicolumn{1}{c}{\textbf{Single}} & \multicolumn{1}{c}{\textbf{Multiple}} & \multicolumn{1}{c}{\textbf{Single}} & \multicolumn{1}{c}{\textbf{Multiple}} & \multicolumn{1}{c}{}                               & \multicolumn{1}{c}{}                                & \multicolumn{1}{c}{}                                           \\ \hline
$\alpha^{(2)}_0$ & $\rho_{\eta_{1}^{(2)}, \eta_{2}^{(2)}}$&                                     &                                       &                                     &                                       &                                     &                                       &                                                    &                                                     &                                                                \\ \hline
0.00             & -0.20                             & 10.14                               & 8.12                                  & 8.57                                & 8.00                                  & 8.64                                & 8.01                                  & MELS                                               & RILM                                                & 1.48                                                           \\ \hline
1.00             & -0.20                             & 13.80                               & 10.29                                 & 11.95                               & 10.10                                 & 11.88                               & 10.04                                 & SPLSME                                             & RILM                                                & 2.43                                                           \\ \hline
2.00             & -0.20                             & 18.80                               & 13.97                                 & 18.38                               & 13.81                                 & 18.88                               & 13.87                                 & MELS                                               & RILM                                                & 1.15                                                           \\ \hline
3.00             & -0.20                             & 36.49                               & 24.79                                 & 38.29                               & 24.62                                 & 37.14                               & 24.52                                 & SPLSME                                             & RILM                                                & 1.09                                                           \\ \hline
0.00             & -0.80                             & 10.92                               & 8.76                                  & 9.23                                & 8.58                                  & 9.24                                & 8.58                                  & MELS                                               & RILM                                                & 2.05                                                           \\ \hline
\end{tabular}
\begin{tablenotes}
\item[a] The best model refers to the model with the smallest error.
\item[b] The worst model refers to the model with the largest error.
\item[c] The difference ratio represents the percentage difference between the model's largest and smallest errors, with the difference normalized by the largest error. For example, the difference ratio "1.68" is calculated by $100\% \times \frac{9.51-9.35}{9.51} \approx 1.68\%$.
\end{tablenotes}
\end{threeparttable}

%% file: bias_0_1_-0.2_1_1_-0.5_0.5_0.1_0.05_0.1.tex
\begin{threeparttable}
\caption{Bias and coverage rates with $\alpha^{(2)}_0 = 0.00$ and $\rho_{\eta_{1}^{(2)}, \eta_{2}^{(2)}} = -0.20$ (or $\gamma^{(2)} = -0.50$ and $\delta^{(2)} = 0.50$) \tnote{a}}
\label{bias_0_1_-0.2_1_1_-0.5_0.5_0.1_0.05_0.1}
\begin{tabular}{lllllllllll}
\hline
\multicolumn{1}{c}{\textbf{}} & \multicolumn{1}{c}{\textbf{}} & \multicolumn{3}{c}{\textbf{RILM}}                          & \multicolumn{3}{c}{\textbf{MELS}}                          & \multicolumn{3}{c}{\textbf{SPLSME}}                          \\ \hline
\textbf{Parameter}            & \textbf{Value}           & \textbf{Estimate} & \textbf{Bias} & \textbf{Coverage} & \textbf{Estimate} & \textbf{Bias} & \textbf{Coverage} & \textbf{Estimate} & \textbf{Bias} & \textbf{Coverage} \\ \hline
$\beta^{(2)}_0$                    & -2.20               & -2.58 & -0.38 & 93 & -1.99 & 0.21  & 98  & -1.75 & 0.45  & 96  \\ \hline

$\beta^{(2)}_1$                         & 2.00                & 1.98  & -0.02 & 88 & 1.99  & -0.01 & 95  & 2.00  & -0.00 & 97  \\ \hline

$\beta^{(2)}_2$                         & 0.02                & 0.04  & 0.02  & 93 & 0.01  & -0.01 & 98  & -0.01 & -0.03 & 96  \\ \hline

$\alpha_0^{(2)}$                   & 0.00                & -0.81 & -0.81 & 12 & 0.22  & 0.22  & 100 & 0.08  & 0.08  & 100 \\ \hline

$\alpha_1^{(2)}$                        & 0.30                & -     & -     & -  & 0.35  & 0.05  & 92  & 0.32  & 0.02  & 97  \\ \hline

$\alpha_2^{(2)}$                         & -0.10               & -     & -     & -  & -0.11 & -0.01 & 100 & -0.10 & -0.00 & 100 \\ \hline

$\gamma^{(2)}$                         & -0.50               & -     & -     & -  & -     & -     & -   & -0.74 & -0.24 & 90  \\ \hline

$\delta^{(2)}$                         & 0.50                & -     & -     & -  & -     & -     & -   & 0.68  & 0.18  & 90  \\ \hline

$\sigma_{v_1^{(2)}}$           & 1.12                & 1.08     & -0.04     & 91  & 1.19  & 0.07  & 96  & 1.26     & 0.14     & 94   \\ \hline

$\sigma_{v_2^{(2)}}$         & 1.12                & -     & -     & -  & 1.17  & 0.05  & 98  & 1.25     & 0.13    & 97  \\ \hline

$\rho_{v_1^{(2)}, v_2^{(2)}}$      & -0.36               & -     & -     & -  & -0.29 & 0.07  & 94  & -0.30     & 0.06     & 96   \\ \hline

$\sigma_{\eta_1^{(2)}}$              & 1.00                & -     & -  &  - & -     & -     & -   & 1.02  & 0.02  & 96  \\ \hline

$\sigma_{\eta_2^{(2)}}$            & 1.00                & -     & -     & -  & -     & -     & -   & 1.00  & -0.00 & 93  \\ \hline

$\rho_{\eta_1^{(2)}, \eta_2^{(2)}}$         & -0.20               & -     & -     & -  & -     & -     & -   & -0.11 & 0.09  & 96  \\ \hline

$\sigma_{\lambda}$              & 1.00                & -     & -     & -  & -     & -     & -   & 0.87  & -0.13 & 95  \\ \hline

$\tau_1$                      & 0.10                & -     & -     & -  & -     & -     & -   & 0.11  & 0.01  & 91  \\ \hline

$\tau_2$                    & 0.05                & -     & -     & -  & -     & -     & -   & -0.00 & -0.05 & 95  \\ \hline

$\tau_3$                    & 0.10                & -     & -     & -  & -     & -     & -   & 0.10  & 0.00  & 99  \\ \hline
\end{tabular}
\begin{tablenotes} 
\item[a] In Subsection \ref{RILM and MELS}, we initially set $\alpha^{(2)} = 0.00$ and $\rho^{(2)}_{\eta_1, \eta_2} = -0.20$, and these parameters are varied in subsequent simulations. Concurrently, the values of $\gamma^{(2)}$ and $\delta^{(2)}$ are consistently maintained at -0.50 and 0.50, respectively. These latter values align with the initial settings for $\gamma^{(2)}$ and $\delta^{(2)}$ in Subsection \ref{MELS and SPLSME}. In Subsection \ref{MELS and SPLSME}, we initially set $\gamma^{(2)} = -0.50$ and $\delta^{(2)} = 0.50$, and these parameters are varied in subsequent simulations. Concurrently, the values of $\alpha^{(2)}$ and $\rho_{\eta^{(2)}_1, \eta^{(2)}_2}$ are consistently maintained at 0.00 and -0.20, respectively. As a result, the table serves a dual purpose: it represents the parameter combination of $\alpha^{(2)} = 0.00$, $\rho_{\eta^{(2)}_1, \eta^{(2)}_2} = -0.20$ from Subsection \ref{RILM and MELS} and the combination $\gamma^{(2)} = -0.50$, $\delta^{(2)} = 0.50$ from Subsection \ref{MELS and SPLSME}.
\end{tablenotes}
\end{threeparttable}

%% file: bias_0_1_-0.8_1_1_-0.5_0.5_0.1_0.05_0.1.tex
\caption{Bias and coverage rates with $\alpha^{(2)}_0 = 0.00$ and $\rho_{\eta_{1}^{(2)}, \eta_{2}^{(2)}} = -0.80$}
\label{bias_0_1_-0.8_1_1_-0.5_0.5_0.1_0.05_0.1}
\begin{tabular}{lllllllllll}
\hline
\multicolumn{1}{c}{\textbf{}} & \multicolumn{1}{c}{\textbf{}} & \multicolumn{3}{c}{\textbf{RILM}}                          & \multicolumn{3}{c}{\textbf{MELS}}                          & \multicolumn{3}{c}{\textbf{SPLSME}}                          \\ \hline
\textbf{Parameter}            & \textbf{Value}           & \textbf{Estimate} & \textbf{Bias} & \textbf{Coverage} & \textbf{Estimate} & \textbf{Bias} & \textbf{Coverage} & \textbf{Estimate} & \textbf{Bias} & \textbf{Coverage} \\ \hline
$\beta^{(2)}_0$                    &-2.20 & -2.85 & -0.65 & 92 & -2.23 & -0.03 & 100 & -2.05 & 0.15  & 99  \\ \hline
$\beta^{(2)}_1$                         & 2.00  & 1.99  & -0.01 & 81 & 2.00  & -0.00 & 95  & 2.00  & 0.00  & 95  \\ \hline
$\beta^{(2)}_2$                         & 0.02  & 0.05  & 0.03  & 93 & 0.02  & -0.00 & 100 & 0.01  & -0.01 & 97  \\ \hline
$\alpha_0^{(2)}$                   & 0.00  & -0.80 & -0.80 & 14 & 0.13  & 0.13  & 100 & -0.03 & -0.03 & 99  \\ \hline
$\alpha_1^{(2)}$                        & 0.30  & -     & -     & -  & 0.33  & 0.03  & 94  & 0.31  & 0.01  & 94  \\ \hline
$\alpha_2^{(2)}$                        & -0.10 & -     & -     & -  & -0.11 & -0.01 & 100 & -0.09 & 0.01  & 100 \\ \hline
$\gamma^{(2)}$                         & -0.50 & -     & -     & -  & -     & -     & -   & -0.78 & -0.28 & 88  \\ \hline
$\delta^{(2)}$                         & 0.50  & -     & -     & -  & -     & -     & -   & 0.76  & 0.26  & 91  \\ \hline
$\sigma_{v_{1}^{(2)}}$           & 1.12  & 1.07  & -0.05 & 94 & 1.14  & 0.02  & 97  & 1.20     & 0.08     & 96   \\ \hline
$\sigma_{v_{2}^{(2)}}$         & 1.12  & -     & -     & -  & 1.12  & 0.00  & 94  & 1.19     & 0.07     & 93   \\ \hline
$\rho_{v_{1}^{(2)}, v_{2}^{(2)}}$      & -0.84 & -     & -     & -  & -0.78 & 0.06  & 99  & -0.78     & 0.06     & 99   \\ \hline
$\sigma_{\eta_{1}^{(2)}}$              & 1.00  & -     & -     & -  & -     & -     & -   & 0.93  & -0.07 & 92  \\ \hline
$\sigma_{\eta_{2}^{(2)}}$            & 1.00  & -     & -     & -  & -     & -     & -   & 0.92  & -0.08 & 93  \\ \hline
$\rho_{\eta_{1}^{(2)}, \eta_{2}^{(2)}}$         & -0.80 & -     & -     & -  & -     & -     & -   & -0.66 & 0.14  & 99  \\ \hline
$\sigma_{\lambda}$              & 1.00  & -     & -     & -  & -     & -     & -   & 0.86  & -0.14 & 95  \\ \hline
$\tau_1$                      & 0.10  & -     & -     & -  & -     & -     & -   & 0.11  & 0.01  & 93  \\ \hline
$\tau_2$                    & 0.05  & -     & -     & -  & -     & -     & -   & 0.00  & -0.05 & 95  \\ \hline
$\tau_3$                    & 0.10  & -     & -     & -  & -     & -     & -   & 0.10  & 0.00  & 99  \\ \hline
\end{tabular}

%% file: error_23.tex
\caption{Imputation errors with different $\gamma^{(2)}$ (the effect coefficient of the random missing effect on the mean of $y$) and $\delta^{(2)}$ (the effect coefficient of the random missing effect on the variance of $y$)}
\label{error_23}
\begin{threeparttable}
\begin{tabular}{lllllllllll}
\hline
\multicolumn{2}{c}{\textbf{Method to impute $x_1$}} & \multicolumn{9}{c}{\textbf{RILM}}                                                                     \\ \hline
\multicolumn{2}{c}{\textbf{Method to impute $y$}}    & \multicolumn{2}{c}{\textbf{RILM}}                                           & \multicolumn{2}{c}{\textbf{MELS}}                                           & \multicolumn{2}{c}{\textbf{SPLSME}}                                         & \multicolumn{1}{c}{\multirow{2}{*}{\textbf{Best}}}\tnote{a} & \multicolumn{1}{c}{\multirow{2}{*}{\textbf{Worst}}}\tnote{b} & \multicolumn{1}{c}{\multirow{2}{*}{\textbf{Difference  (\%)}}}\tnote{c} \\ \cline{1-8}
\multicolumn{2}{c}{\textbf{Imputation method}}       & \multicolumn{1}{c}{\textbf{Single}} & \multicolumn{1}{c}{\textbf{Multiple}} & \multicolumn{1}{c}{\textbf{Single}} & \multicolumn{1}{c}{\textbf{Multiple}} & \multicolumn{1}{c}{\textbf{Single}} & \multicolumn{1}{c}{\textbf{Multiple}} & \multicolumn{1}{c}{}                               & \multicolumn{1}{c}{}                                & \multicolumn{1}{c}{}                                           \\ \hline
$\gamma^{(2)}$           & $\delta^{(2)}$           &                                     &                                       &                                     &                                       &                                     &                                       &                                                    &                                                     &                                                                \\ \hline
-0.50                    & 0.50                     & 11.11                               & 9.51                                  & 9.98                                & 9.35                                  & 10.10                               & 9.40                                  & MELS                                               & RILM                                                & 1.68                                                           \\ \hline
-1.00                    & 0.50                     & 11.06                               & 9.01                                  & 9.70                                & 9.04                                  & 9.87                                & 9.10                                  & SPLSME                                             & RILM                                                & 0.99                                                           \\ \hline
-1.50                    & 0.50                     & 14.46                               & 8.28                                  & 8.71                                & 8.01                                  & 8.73                                & 8.06                                  & MELS                                               & RILM                                                & 3.26                                                           \\ \hline
-0.50                    & 0.75                     & 12.44                               & 10.21                                 & 11.01                               & 10.06                                 & 11.03                               & 10.09                                 & MELS                                               & RILM                                                & 1.47                                                           \\ \hline
-0.50                    & 1.00                     & 13.22                               & 10.76                                 & 11.78                               & 10.63                                 & 11.81                               & 10.67                                 & MELS                                               & RILM                                                & 1.21                                                           \\ \hline
\multicolumn{2}{c}{\textbf{Method to impute $x_1$}} & \multicolumn{9}{c}{\textbf{MELS}}                                                                                                                                                                                                                                                                                                                                                                                   \\ \hline
\multicolumn{2}{c}{\textbf{Method to impute $y$}}   & \multicolumn{2}{c}{\textbf{RILM}}                                           & \multicolumn{2}{c}{\textbf{MELS}}                                           & \multicolumn{2}{c}{\textbf{SPLSME}}                                         & \multicolumn{1}{c}{\multirow{2}{*}{\textbf{Best}}} & \multicolumn{1}{c}{\multirow{2}{*}{\textbf{Worst}}} & \multicolumn{1}{c}{\multirow{2}{*}{\textbf{Difference  (\%)}}} \\ \cline{1-8}
\multicolumn{2}{c}{\textbf{Imputation method}}      & \multicolumn{1}{c}{\textbf{Single}} & \multicolumn{1}{c}{\textbf{Multiple}} & \multicolumn{1}{c}{\textbf{Single}} & \multicolumn{1}{c}{\textbf{Multiple}} & \multicolumn{1}{c}{\textbf{Single}} & \multicolumn{1}{c}{\textbf{Multiple}} & \multicolumn{1}{c}{}                               & \multicolumn{1}{c}{}                                & \multicolumn{1}{c}{}                                           \\ \hline
$\gamma^{(2)}$           & $\delta^{(2)}$           &                                     &                                       &                                     &                                       &                                     &                                       &                                                    &                                                     &                                                                \\ \hline
-0.50                    & 0.50                     & 10.37                               & 8.40                                  & 8.94                                & 8.33                                  & 9.00                                & 8.36                                  & MELS                                               & RILM                                                & 0.83                                                           \\ \hline
-1.00                    & 0.50                     & 10.41                               & 8.01                                  & 8.53                                & 7.90                                  & 8.75                                & 7.96                                  & MELS                                               & RILM                                                & 1.37                                                           \\ \hline
-1.50                    & 0.50                     & 13.13                               & 7.44                                  & 7.58                                & 6.87                                  & 7.55                                & 6.91                                  & MELS                                               & RILM                                                & 7.66                                                           \\ \hline
-0.50                    & 0.75                     & 11.32                               & 9.18                                  & 9.99                                & 9.07                                  & 9.99                                & 9.10                                  & MELS                                               & RILM                                                & 1.20                                                           \\ \hline
-0.50                    & 1.00                     & 11.85                               & 9.44                                  & 10.56                               & 9.33                                  & 10.51                               & 9.38                                  & MELS                                               & RILM                                                & 1.17                                                           \\ \hline
\multicolumn{2}{c}{\textbf{Method to impute $x_1$}} & \multicolumn{9}{c}{\textbf{SPLSME}}                                                                                                                                                                                                                                                                                                                                                                                   \\ \hline
\multicolumn{2}{c}{\textbf{Method to impute $y$}}   & \multicolumn{2}{c}{\textbf{RILM}}                                           & \multicolumn{2}{c}{\textbf{MELS}}                                           & \multicolumn{2}{c}{\textbf{SPLSME}}                                         & \multicolumn{1}{c}{\multirow{2}{*}{\textbf{Best}}} & \multicolumn{1}{c}{\multirow{2}{*}{\textbf{Worst}}} & \multicolumn{1}{c}{\multirow{2}{*}{\textbf{Difference  (\%)}}} \\ \cline{1-8}
\multicolumn{2}{c}{\textbf{Imputation method}}      & \multicolumn{1}{c}{\textbf{Single}} & \multicolumn{1}{c}{\textbf{Multiple}} & \multicolumn{1}{c}{\textbf{Single}} & \multicolumn{1}{c}{\textbf{Multiple}} & \multicolumn{1}{c}{\textbf{Single}} & \multicolumn{1}{c}{\textbf{Multiple}} & \multicolumn{1}{c}{}                               & \multicolumn{1}{c}{}                                & \multicolumn{1}{c}{}                                           \\ \hline
$\gamma^{(2)}$           & $\delta^{(2)}$           &                                     &                                       &                                     &                                       &                                     &                                       &                                                    &                                                     &                                                                \\ \hline
-0.50                    & 0.50                     & 10.14                               & 8.12                                  & 8.57                                & 8.00                                  & 8.64                                & 8.01                                  & MELS                                               & RILM                                                & 1.50                                                           \\ \hline
-1.00                    & 0.50                     & 10.29                               & 7.85                                  & 8.33                                & 7.66                                  & 8.41                                & 7.64                                  & SPLSME                                             & RILM                                                & 2.68                                                           \\ \hline
-1.50                    & 0.50                     & 13.08                               & 7.41                                  & 7.41                                & 6.67                                  & 7.30                                & 6.63                                  & SPLSME                                             & RILM                                                & 10.53                                                          \\ \hline
-0.50                    & 0.75                     & 11.08                               & 8.93                                  & 9.62                                & 8.79                                  & 9.63                                & 8.77                                  & SPLSME                                             & RILM                                                & 1.79                                                           \\ \hline
-0.50                    & 1.00                     & 11.56                               & 9.20                                  & 10.24                               & 9.04                                  & 10.25                               & 9.06                                  & MELS                                               & RILM                                                & 1.74                                                           \\ \hline
\end{tabular}
\begin{tablenotes} 
\item[a] The best model refers to the model with the smallest error.
\item[b] The worst model refers to the model with the largest error.
\item[c] The difference represents the percentage difference between the model's largest and smallest errors, with the difference normalized by the largest error. For example, the difference "1.68" is calculated by $100\% \times \frac{9.51-9.35}{9.51} \approx 1.68\%$.
\end{tablenotes}
\end{threeparttable}

%% file: bias_0_1_-0.2_1_1_-0.5_1_0.1_0.05_0.1.tex
\caption{Bias and coverage rates with $\gamma^{(2)} = -0.50$ and $\delta^{(2)} = 1.00$}
\label{bias_0_1_-0.2_1_1_-0.5_1_0.1_0.05_0.1}
\begin{tabular}{lllllllllll}
\hline
\multicolumn{1}{c}{\textbf{}} & \multicolumn{1}{c}{\textbf{}} & \multicolumn{3}{c}{\textbf{RILM}}                          & \multicolumn{3}{c}{\textbf{MELS}}                          & \multicolumn{3}{c}{\textbf{SPLSME}}                          \\ \hline
\textbf{Parameter}            & \textbf{Value}           & \textbf{Estimate} & \textbf{Bias} & \textbf{Coverage} & \textbf{Estimate} & \textbf{Bias} & \textbf{Coverage} & \textbf{Estimate} & \textbf{Bias} & \textbf{Coverage} \\ \hline
$\beta^{(2)}_0$                    & -2.20 & -2.74 & -0.54 & 96 & -2.04 & 0.16  & 98  & -1.92 & 0.28  & 95  \\ \hline

$\beta^{(2)}_1$                         & 2.00  & 1.98  & -0.02 & 80 & 2.00  & 0.00  & 94  & 2.00  & 0.00  & 90  \\ \hline

$\beta^{(2)}_2$                         & 0.02  & 0.04  & 0.02  & 94 & 0.01  & -0.01 & 97  & 0.00  & -0.02 & 94  \\ \hline

$\alpha_0^{(2)}$                   & 0.00  & -0.39 & -0.39 & 42 & 0.35  & 0.35  & 100 & 0.12  & 0.12  & 100 \\ \hline

$\alpha_1^{(2)}$                        & 0.30  & -     & -     & -  & 0.36  & 0.06  & 86  & 0.32  & 0.02  & 93  \\ \hline

$\alpha_2^{(2)}$                         & -0.10 & -     & -     & -  & -0.11 & -0.01 & 100 & -0.10 & 0.00  & 98  \\ \hline

$\gamma^{(2)}$                         & -0.50 & -     & -     & -  & -     & -     & -   & -0.71 & -0.21 & 96  \\ \hline

$\delta^{(2)}$                         & 1.00  & -     & -     & -  & -     & -     & -   & 1.29  & 0.29  & 93  \\ \hline

$\sigma_{v_{1, i}^{(2)}}$           & 1.12  & 1.07  & -0.05 & 92 & 1.18  & 0.06  & 96  & 1.24     & 0.12     & 96   \\ \hline

$\sigma_{v_{2, i}^{(2)}}$         & 1.41  & -     & -     & -  & 1.42  & 0.00  & 98  & 1.49     & 0.08     & 99   \\ \hline

$\rho_{v_{1, i}^{(2)}, v_{2, i}^{(2)}}$      & -0.44 & -     & -     & -  & -0.40 & 0.04  & 96  & -0.41     & 0.03     & 95   \\ \hline

$\sigma_{\eta_{1, i}^{(2)}}$              & 1.00  & -     & -     & -  & -     & -     & -   & 1.02  & 0.02  & 98  \\ \hline

$\sigma_{\eta_{2, i}^{(2)}}$            & 1.00  & -     & -     & -  & -     & -     & -   & 0.95  & -0.05 & 96  \\ \hline

$\rho_{\eta_{1, i}^{(2)}, \eta_{2, i}^{(2)}}$         & -0.20 & -     & -     & -  & -     & -     & -   & -0.12 & 0.08  & 98  \\ \hline

$\sigma_{\lambda_i}$              & 1.00  & -     & -     & -  & -     & -     & -   & 1.49  & 0.49 & 95  \\ \hline

$\tau_1$                      & 0.10  & -     & -     & -  & -     & -     & -   & 0.11  & 0.01  & 92  \\ \hline

$\tau_2$                    & 0.05  & -     & -     & -  & -     & -     & -   & 0.00  & -0.05 & 95  \\ \hline

$\tau_3$                    & 0.10  & -     & -     & -  & -     & -     & -   & 0.10  & 0.00  & 98  \\ \hline
\end{tabular}

%% file: example_PA.tex
\begin{threeparttable}
\caption{Parameter estimates and intervals for PA}
\label{example_PA}
\begin{tabular}{lllllllllll}
\hline
\multicolumn{1}{c}{\textbf{}} & \multicolumn{2}{c}{\textbf{RILM}}                          & \multicolumn{2}{c}{\textbf{MELS}}                          & \multicolumn{2}{c}{\textbf{SPLSME}}                          \\ \hline
\textbf{Parameter}           & \textbf{Estimate} & \textbf{95\% Confidence Interval} & \textbf{Estimate} & \textbf{95\% Credible Interval\tnote{a}} & \textbf{Estimate} & \textbf{95\% Credible Interval} \\ \hline
$\beta_{inter}$         & 8.05     & (4.11, 11.99)    & 8.00      & (3.22, 12.84)    & 9.52     & (4.85, 13.93)     \\ \hline
$\beta_{day}$           & -0.04    & (-0.07, -0.02)   & -0.02     & (-0.04, +0.00\tnote{b})    & -0.02    & (-0.04, +0.00)     \\ \hline
$\beta_{sex}$           & -0.48    & (-1.14, 0.18)    & -0.50     & (-1.36, 0.34)    & -0.52    & (-1.29, 0.22)     \\ \hline
$\beta_{age}$           & -0.04    & (-0.10, 0.01)    & -0.04     & (-0.11, 0.02)    & -0.06    & (-0.13, +0.00)     \\ \hline
$\beta_{edu2}$          & +0.00     & (-0.77, 0.77)    & 0.01      & (-0.95, 1.04)    & -0.22   & (-1.10, 0.68)     \\ \hline
$\beta_{edu3}$          & 0.59     & (-0.16, 1.34)    & 0.62      & (-0.28, 1.53)    & 0.64     & (-0.22, 1.48)     \\ \hline
$\beta_{year}$          & 0.04     & (-0.05, 0.14)    & 0.05      & (-0.07, 0.17)    & 0.05     & (-0.06, 0.15)     \\ \hline
$\beta_{CDR2}$          & -0.19    & (-0.96, 0.58)    & -0.14     & (-1.14, 0.87)    & -0.21    & (-1.08, 0.69)     \\ \hline
$\beta_{CDR3}$          & 1.02     & (-0.90, 2.94)    & 1.10      & (-1.30, 3.46)    & 0.20     & (-2.28, 2.58)     \\ \hline
$\alpha_{inter}$        & -0.38    & (-0.68, -0.09)   & -2.66     & (-6.56, 0.90)    & -3.20    & (-7.10, 0.38)     \\ \hline
$\alpha_{day}$          & -        & -                & -0.12     & (-0.17, -0.06)   & -0.12    & (-0.17, -0.07)    \\ \hline
$\alpha_{sex}$          & -        & -                & 0.16      & (-0.47, 0.81)    & 0.19     & (-0.46, 0.90)     \\ \hline
$\alpha_{age}$          & -        & -                & 0.04      & (-0.01, 0.09)    & 0.05     & (-0.01, 0.10)     \\ \hline
$\alpha_{edu2}$         & -        & -                & 0.23      & (-0.61, 1.02)    & 0.34     & (-0.38, 1.05)     \\ \hline
$\alpha_{edu3}$         & -        & -                & -0.31     & (-1.01, 0.38)    & -0.33    & (-1.04, 0.41)     \\ \hline
$\alpha_{year}$         & -        & -                & -0.03     & (-0.13, 0.08)    & -0.02    & (-0.12, 0.07)     \\ \hline
$\alpha_{CDR2}$         & -        & -                & -0.27     & (-1.06, 0.54)    & -0.25    & (-1.06, 0.58)     \\ \hline
$\alpha_{CDR3}$         & -        & -                & 0.16      & (-1.71, 2.09)    & 0.49     & (-1.42, 2.28)     \\ \hline
$\tau_{inter}$          & -        & -                & -         & -                & -0.97    & (-1.40, -0.55)    \\ \hline
$\tau_{day}$            & -        & -                & -         & -                & 0.03    & (-0.04, 0.11)     \\ \hline
$\tau_{{beep}_{3, 4}}$  & -        & -                & -         & -                & -0.85    & (-1.22,  -0.47)   \\ \hline
$\tau_{{beep}_{5, 6}}$  & -        & -                & -         & -                & -0.63    & (-0.98, -0.28)    \\ \hline
$\tau_{{beep}_{7, 8}}$  & -        & -                & -         & -                & -0.84    & (-1.22, -0.47)    \\ \hline
$\tau_{{beep}_{9, 10}}$ & -        & -                & -         & -                & -1.22    & (-1.61, -0.82)    \\ \hline
$\gamma$                & -        & -                & -         & -                & -0.96    & (-1.87, -0.17)    \\ \hline
$\delta$                & -        & -                & -         & -                & 0.34     & (-0.34, 1.11)     \\ \hline
$\sigma_{v_1}$          & 0.85     & (0.57, 0.97)     & 0.94      & (0.67, 1.31)     & 0.97        & (0.70, 1.38)                 \\ \hline
$\sigma_{v_2}$          & -        & -                & 0.71      & (0.51, 1.03)     & 0.75        & (0.53, 1.06)                 \\ \hline
$\sigma_{\eta_1}$       & -        & -                & -         & -                & 0.78     & (0.51, 1.16)      \\ \hline
$\sigma_{\eta_2}$       & -        & -                & -         & -                & 0.70     & (0.49, 0.98)      \\ \hline
$\sigma_{\lambda}$       & -        & -                & -         & -                & 0.57     & (0.37, 0.84)      \\ \hline
$\rho_{v_1, v_2}$       & -        & -                & -0.43     & (-0.74, -0.04)   & -0.44        & (-0.75, -0.02)                 \\ \hline
$\rho_{\eta_1, \eta_2}$ & -        & -                & -         & -                & -0.36    & (-0.73, 0.12)     \\ \hline
BIC                     & \multicolumn{2}{l}{4520.60} & \multicolumn{2}{l}{3353.12}  & \multicolumn{2}{l}{3422.19}  \\ \hline
ELPD                    & \multicolumn{2}{l}{-}       & \multicolumn{2}{l}{-1673.73} & \multicolumn{2}{l}{-1672.27} \\ \hline
\end{tabular}
\begin{tablenotes} 
\item[a] MELS and SPLSME are estimated by the MCMC approach instead of MLE. Therefore, we present the credible intervals by determining the lower 2.5\% and upper 97.5\% quantiles for the sample set of each estimate.
\item[b] The values displayed in the table have been rounded to two decimal places. A value of "+0.00" indicates that the original number, before rounding, was greater than 0. Conversely, "-0.00" signifies that the original number was less than 0 before rounding.
\end{tablenotes}
\end{threeparttable}

%% file: example_stress.tex
\begin{threeparttable}
\caption{Parameter estimates and intervals for stress}
\label{example_stress}
\begin{tabular}{lllllllllll}
\hline
\multicolumn{1}{c}{\textbf{}} & \multicolumn{2}{c}{\textbf{RILM}}                          & \multicolumn{2}{c}{\textbf{MELS}}                          & \multicolumn{2}{c}{\textbf{SPLSME}}                          \\ \hline
\textbf{Parameter}           & \textbf{Estimate} & \textbf{95\% Confidence Interval} & \textbf{Estimate} & \textbf{95\% Credible Interval\tnote{a}} & \textbf{Estimate} & \textbf{95\% Credible Interval} \\ \hline
$\beta_{inter}$         & 4.99  & (2.63, 7.37)  & 4.37  & (1.19, 7.43)   & 4.47  & (1.18, 7.83)   \\ \hline
$\beta_{PA}$            & -0.47 & (-0.54, -0.41) & -0.42 & (-0.48, -0.36) & -0.42 & (-0.48, -0.35) \\ \hline
$\beta_{sex}$           & -0.07 & (-0.46, 0.32) & -0.03 & (-0.54, 0.47)  & -0.01 & (-0.54, 0.51)  \\ \hline
$\beta_{age}$           & 0.00  & (-0.03, 0.04) & 0.01  & (-0.03, 0.05)  & 0.01  & (-0.04, 0.05)  \\ \hline
$\beta_{edu2}$          & -0.11 & (-0.57, 0.34) & -0.12 & (-0.75, 0.46)  & -0.12 & (-0.78, 0.48)  \\ \hline
$\beta_{edu3}$          & -0.12 & (-0.55, 0.33) & -0.14 & (-0.71, 0.43)  & -0.13 & (-0.70, 0.44)  \\ \hline
$\beta_{year}$          & 0.00  & (-0.05, 0.06) & -0.00 & (-0.07, 0.07)  & -0.00 & (-0.07, 0.07)  \\ \hline
$\beta_{CDR2}$          & -0.07 & (-0.52, 0.39) & -0.05 & (-0.60, 0.52)  & -0.06 & (-0.65, 0.57)  \\ \hline
$\beta_{CDR3}$          & -0.40 & (-1.52, 0.73) & -0.41 & (-1.92, 1.10)  & -0.48 & (-2.14, 1.13)  \\ \hline
$\alpha_{inter}$        & 0.17  & (-0.09, 0.42) & 3.06  & (0.53, 5.59)   & 3.87  & (1.28, 6.41)   \\ \hline
$\alpha_{PA}$           & -     & -             & -0.15 & (-0.23, -0.06) & -0.16 & (-0.25, -0.08) \\ \hline
$\alpha_{sex}$          & -     & -             & -0.17 & (-0.60, 0.25)  & -0.18 & (-0.59, 0.22)  \\ \hline
$\alpha_{age}$          & -     & -             & -0.04 & (-0.07, -0.00\tnote{b})  & -0.04 & (-0.08, -0.01) \\ \hline
$\alpha_{edu2}$         & -     & -             & 0.15  & (-0.29, 0.63)  & 0.05  & (-0.38, 0.51)  \\ \hline
$\alpha_{edu3}$         & -     & -             & 0.07  & (-0.40, 0.53)  & 0.06  & (-0.38, 0.49)  \\ \hline
$\alpha_{year}$         & -     & -             & 0.04  & (-0.02, 0.10)  & 0.04  & (-0.02, 0.10)  \\ \hline
$\alpha_{CDR2}$         & -     & -             & -0.09 & (-0.60, 0.38)  & -0.13 & (-0.58, 0.34)  \\ \hline
$\alpha_{CDR3}$         & -     & -             & -0.50 & (-1.71, 0.78)  & -0.90 & (-2.11, 0.30)  \\ \hline
$\tau_{inter}$          & -     & -             & -     & -              & -0.96 & (-1.34, -0.58) \\ \hline
$\tau_{day}$            & -     & -             & -     & -              & 0.03  & (-0.04, 0.10)  \\ \hline
$\tau_{{beep}_{3, 4}}$  & -     & -             & -     & -              & -0.85 & (-1.23, -0.48) \\ \hline
$\tau_{{beep}_{5, 6}}$  & -     & -             & -     & -              & -0.63 & (-0.96, -0.29) \\ \hline
$\tau_{{beep}_{7, 8}}$  & -     & -             & -     & -              & -0.85 & (-1.21, -0.48) \\ \hline
$\tau_{{beep}_{9, 10}}$ & -     & -             & -     & -              & -1.22 & (-1.60, -0.86) \\ \hline
$\gamma$                & -     & -             & -     & -              & -0.04 & (-0.59, 0.54)  \\ \hline
$\delta$                & -     & -             & -     & -              & -0.44 & (-0.94, -0.04) \\ \hline
$\sigma_{v_1}$          & 0.48  & (0.31, 0.55)  & 0.55  & (0.39, 0.77)   & 0.57     & (0.41, 0.80)              \\ \hline
$\sigma_{v_2}$          & -     & -             & 0.43  & (0.29, 0.62)   & 0.45     & (0.30, 0.67)              \\ \hline
$\sigma_{\eta_1}$       & -     & -             & -     & -              & 0.54  & (0.39, 0.77)   \\ \hline
$\sigma_{\eta_2}$       & -     & -             & -     & -              & 0.36  & (0.17, 0.58)   \\ \hline
$\sigma_{\lambda}$       & -     & -             & -     & -              & 0.57  & (0.35, 0.86)   \\ \hline
$\rho_{v_1, v_2}$       & -     & -             & -0.17 & (-0.59, 0.29)  & -0.17     & (-0.60, 0.31)              \\ \hline
$\rho_{\eta_1, \eta_2}$ & -     & -             & -     & -              & -0.24 & (-0.70, 0.27)  \\ \hline
BIC                     & \multicolumn{2}{l}{4514.65} & \multicolumn{2}{l}{4323.17}  & \multicolumn{2}{l}{4391.14}  \\ \hline
ELPD                    & \multicolumn{2}{l}{-}       & \multicolumn{2}{l}{-2142.51} & \multicolumn{2}{l}{-2142.91} \\ \hline
\end{tabular}
\begin{tablenotes} 
\item[a] MELS and SPLSME are estimated by the MCMC approach instead of MLE. Therefore, we present the credible intervals by determining the lower 2.5\% and upper 97.5\% quantiles for the sample set of each estimate.
\item[b] The values displayed in the table have been rounded to two decimal places. A value of "+0.00" indicates that the original number, before rounding, was greater than 0. Conversely, "-0.00" signifies that the original number was less than 0 before rounding.
\end{tablenotes}
\end{threeparttable}

%% file: bias_1_1_-0.2_1_1_-0.5_0.5_0.1_0.05_0.1.tex
\caption{Bias and coverage rates with $\alpha^{(2)}_0 = 1.00$ and $\rho_{\eta_{1}^{(2)}, \eta_{2}^{(2)}} = -0.20$}
\label{bias_1_1_-0.2_1_1_-0.5_0.5_0.1_0.05_0.1}
\begin{tabular}{lllllllllll}
\hline
\multicolumn{1}{c}{\textbf{}} & \multicolumn{1}{c}{\textbf{}} & \multicolumn{3}{c}{\textbf{RILM}}                          & \multicolumn{3}{c}{\textbf{MELS}}                          & \multicolumn{3}{c}{\textbf{SPLSME}}                          \\ \hline
\textbf{Parameter}            & \textbf{Value}           & \textbf{Estimate} & \textbf{Bias} & \textbf{Coverage} & \textbf{Estimate} & \textbf{Bias} & \textbf{Coverage} & \textbf{Estimate} & \textbf{Bias} & \textbf{Coverage} \\ \hline
$\beta^{(2)}_0$              &     -2.20 & -2.81 & -0.61 & 93 & -2.02 & 0.18  & 99  & -1.83 & 0.37  & 98 \\ \hline

$\beta^{(2)}_1$                         & 2.00  & 1.95  & -0.05 & 79 & 1.99  & -0.01 & 94  & 2.00  & -0.00 & 95 \\ \hline

$\beta^{(2)}_2$                         & 0.02  & 0.05  & 0.03  & 89 & 0.01  & -0.01 & 100 & -0.00 & -0.02 & 99 \\ \hline

$\alpha_0^{(2)}$                  & 1.00  & 0.19  & -0.81 & 6  & 0.57  & -0.43 & 100 & 0.41  & -0.59 & 98 \\ \hline

$\alpha_1^{(2)}$                       & 0.30  & -     & -     & -  & 0.32  & 0.02  & 95  & 0.29  & -0.01 & 95 \\ \hline

$\alpha_2^{(2)}$                        & -0.10 & -     & -     & -  & -0.08 & 0.02  & 99  & -0.07 & 0.03  & 95 \\ \hline

$\gamma^{(2)}$                         & -0.50 & -     & -     & -  & -     & -     & -   & -0.73 & -0.23 & 95 \\ \hline

$\delta^{(2)}$                         & 0.50  & -     & -     & -  & -     & -     & -   & 0.76  & 0.26  & 92 \\ \hline

$\sigma_{v_{1}^{(2)}}$           & 1.12  & 1.08  & -0.04    & 92 & 1.19  & 0.07     & 97 & 1.26     & 0.14     & 95  \\ \hline

$\sigma_{v_{2}^{(2)}}$         & 1.12  & -     & -     & -     & 1.21  & 0.09  & 100   & 1.29     & 0.17     & 98  \\ \hline

$\rho_{v_{1}^{(2)}, v_{2}^{(2)}}$      & -0.36 & -     & -     & -     & -0.30 & 0.06  & 97    & -0.32     & 0.04     & 98  \\ \hline

$\sigma_{\eta_{1}^{(2)}}$             & 1.00  & -     & -     & -     & -     & -     & -     & 1.02  & 0.02  & 93 \\ \hline

$\sigma_{\eta_{2}^{(2)}}$            & 1.00  & -     & -     & -     & -     & -     & -     & 1.02  & 0.02  & 97 \\ \hline

$\rho_{\eta_{1}^{(2)}, \eta_{2}^{(2)}}$         & -0.20 & -     & -     & -     & -     & -     & -     & -0.10 & 0.10  & 95 \\ \hline

$\sigma_{\lambda}$              & 1.00  & -     & -     & -     & -     & -     & -     & 0.87  & -0.13 & 94 \\ \hline

$\tau_1$                      & 0.10  & -     & -     & -     & -     & -     & -     & 0.11  & 0.01  & 93 \\ \hline

$\tau_2$                    & 0.05  & -     & -     & -     & -     & -     & -     & 0.00  & -0.05 & 95 \\ \hline

$\tau_3$                    & 0.10  & -     & -     & -     & -     & -     & -     & 0.10  & 0.00  & 99 \\ \hline
\end{tabular}

%% file: bias_2_1_-0.2_1_1_-0.5_0.5_0.1_0.05_0.1.tex
\caption{Bias and coverage rates with $\alpha^{(2)}_0 = 2.00$ and $\rho_{\eta_{1}^{(2)}, \eta_{2}^{(2)}} = -0.20$}
\label{bias_2_1_-0.2_1_1_-0.5_0.5_0.1_0.05_0.1}
\begin{tabular}{lllllllllll}
\hline
\multicolumn{1}{c}{\textbf{}} & \multicolumn{1}{c}{\textbf{}} & \multicolumn{3}{c}{\textbf{RILM}}                          & \multicolumn{3}{c}{\textbf{MELS}}                          & \multicolumn{3}{c}{\textbf{SPLSME}}                          \\ \hline
\textbf{Parameter}            & \textbf{Value}           & \textbf{Estimate} & \textbf{Bias} & \textbf{Coverage} & \textbf{Estimate} & \textbf{Bias} & \textbf{Coverage} & \textbf{Estimate} & \textbf{Bias} & \textbf{Coverage} \\ \hline
$\beta^{(2)}_0$                    & -2.20 & -2.79 & -0.59 & 86 & -1.84 & 0.36  & 98 & -1.66 & 0.54  & 95 \\ \hline
$\beta^{(2)}_1$                         & 2.00  & 1.94  & -0.06 & 77 & 1.98  & -0.02 & 97 & 2.00  & -0.00 & 97 \\ \hline
$\beta^{(2)}_2$                         & 0.02  & 0.05  & 0.03  & 88 & 0.00  & -0.02 & 99 & -0.01 & -0.03 & 95 \\ \hline
$\alpha_0^{(2)}$                   & 2.00  & 1.18  & -0.82 & 4  & 1.07  & -0.93 & 89 & 0.93  & -1.07 & 81 \\ \hline
$\alpha_1^{(2)}$                        & 0.30  & -     & -     & -  & 0.32  & 0.02  & 92 & 0.29  & -0.01 & 92 \\ \hline
$\alpha_2^{(2)}$                        & -0.10 & -     & -     & -  & -0.05 & 0.05  & 89 & -0.04 & 0.06  & 82 \\ \hline
$\gamma^{(2)}$                         & -0.50 & -     & -     & -  & -     & -     & -  & -0.77 & -0.27 & 93 \\ \hline
$\delta^{(2)}$                         & 0.50  & -     & -     & -  & -     & -     & -  & 0.76  & 0.26  & 91 \\ \hline
$\sigma_{v_{1}^{(2)}}$           & 1.12  & 1.10  & -0.01 & 94 & 1.24  & 0.12  & 98 & 1.31     & 0.20     & 95  \\ \hline
$\sigma_{v_{2}^{(2)}}$         & 1.12  & -     & -     & -  & 1.21  & 0.09  & 94 & 1.29     & 0.17     & 92  \\ \hline
$\rho_{v_{1}^{(2)}, v_{2}^{(2)}}$      & -0.36 & -     & -     & -  & -0.28 & 0.08  & 96 & -0.31     & 0.05     & 99  \\ \hline
$\sigma_{\eta_{1}^{(2)}}$              & 1.00  & -     & -     & -  & -     & -     & -  & 1.05  & 0.05  & 97 \\ \hline
$\sigma_{\eta_{2}^{(2)}}$            & 1.00  & -     & -     & -  & -     & -     & -  & 1.04  & 0.04  & 96 \\ \hline
$\rho_{\eta_{1}^{(2)}, \eta_{2}^{(2)}}$         & -0.20 & -     & -     & -  & -     & -     & -  & -0.12 & 0.08  & 98 \\ \hline
$\sigma_{\lambda_i}$              & 1.00  & -     & -     & -  & -     & -     & -  & 0.86  & -0.14 & 95 \\ \hline
$\tau_1$                      & 0.10  & -     & -     & -  & -     & -     & -  & 0.11  & 0.01  & 94 \\ \hline
$\tau_2$                    & 0.05  & -     & -     & -  & -     & -     & -  & 0.00  & -0.05 & 95 \\ \hline
$\tau_3$                    & 0.10  & -     & -     & -  & -     & -     & -  & 0.10  & 0.00  & 99 \\ \hline
\end{tabular}

%% file: bias_3_1_-0.2_1_1_-0.5_0.5_0.1_0.05_0.1.tex
\caption{Bias and coverage rates with $\alpha^{(2)}_0 = 3.00$ and $\rho_{\eta_{1}^{(2)}, \eta_{2}^{(2)}} = -0.20$}
\label{bias_3_1_-0.2_1_1_-0.5_0.5_0.1_0.05_0.1}
\begin{tabular}{lllllllllll}
\hline
\multicolumn{1}{c}{\textbf{}} & \multicolumn{1}{c}{\textbf{}} & \multicolumn{3}{c}{\textbf{RILM}}                          & \multicolumn{3}{c}{\textbf{MELS}}                          & \multicolumn{3}{c}{\textbf{SPLSME}}                          \\ \hline
\textbf{Parameter}            & \textbf{Value}           & \textbf{Estimate} & \textbf{Bias} & \textbf{Coverage} & \textbf{Estimate} & \textbf{Bias} & \textbf{Coverage} & \textbf{Estimate} & \textbf{Bias} & \textbf{Coverage} \\ \hline
$\beta^{(2)}_0$                    & -2.20 & -2.91 & -0.71 & 91 & -1.77 & 0.43  & 99  & -1.59 & 0.61  & 97 \\ \hline
$\beta^{(2)}_1$                         & 2.00  & 1.91  & -0.09 & 78 & 1.97  & -0.03 & 96  & 2.01  & 0.01  & 94 \\ \hline
$\beta^{(2)}_2$                         & 0.02  & 0.06  & 0.04  & 87 & 0.00  & -0.02 & 100 & -0.01 & -0.03 & 97 \\ \hline
$\alpha_0^{(2)}$                   & 3.00  & 2.15  & -0.85 & 4  & 1.33  & -1.67 & 43  & 1.23  & -1.77 & 31 \\ \hline
$\alpha_1^{(2)}$                        & 0.30  & -     & -     & -  & 0.31  & 0.01  & 94  & 0.28  & -0.02 & 95 \\ \hline
$\alpha_2^{(2)}$                        & -0.10 & -     & -     & -  & -0.02 & 0.08  & 52  & -0.01 & 0.09  & 43 \\ \hline
$\gamma^{(2)}$                         & -0.50 & -     & -     & -  & -     & -     & -   & -0.71 & -0.21 & 94 \\ \hline
$\delta^{(2)}$                         & 0.50  & -     & -     & -  & -     & -     & -   & 0.79  & 0.29  & 94 \\ \hline
$\sigma_{v_{1}^{(2)}}$           & 1.12  & 1.07  & -0.05 & 89 & 1.19  & 0.07  & 95  & 1.27     & 0.15     & 94  \\ \hline
$\sigma_{v_{2}^{(2)}}$         & 1.12  & -     & -     & -  & 1.35  & 0.23  & 85  & 1.42     & 0.31     & 81  \\ \hline
$\rho_{v_{1}^{(2)}, v_{2}^{(2)}}$      & -0.36 & -     & -     & -  & -0.28 & 0.08  & 97  & -0.31     & 0.05     & 99  \\ \hline
$\sigma_{\eta_{1}^{(2)}}$              & 1.00  & -     & -     & -  & -     & -     & -   & 0.99  & -0.01 & 97 \\ \hline
$\sigma_{\eta_{2}^{(2)}}$            & 1.00  & -     & -     & -  & -     & -     & -   & 1.16  & 0.16  & 93 \\ \hline
$\rho_{\eta_{1}^{(2)}, \eta_{2}^{(2)}}$         & -0.20 & -     & -     & -  & -     & -     & -   & -0.12 & 0.08  & 95 \\ \hline
$\sigma_{\lambda_i}$              & 1.00  & -     & -     & -  & -     & -     & -   & 0.88  & -0.12 & 95 \\ \hline
$\tau_1$                      & 0.10  & -     & -     & -  & -     & -     & -   & 0.11  & 0.01  & 94 \\ \hline
$\tau_2$                    & 0.05  & -     & -     & -  & -     & -     & -   & 0.00  & -0.05 & 94 \\ \hline
$\tau_3$                    & 0.10  & -     & -     & -  & -     & -     & -   & 0.10  & 0.00  & 99 \\ \hline
\end{tabular}

%% file: bias_0_1_-0.2_1_1_-1_0.5_0.1_0.05_0.1.tex
\caption{Bias and coverage rates with $\gamma^{(2)} = -1.00$ and $\delta^{(2)} = 0.50$}
\label{bias_0_1_-0.2_1_1_-1_0.5_0.1_0.05_0.1}
\begin{tabular}{lllllllllll}
\hline
\multicolumn{1}{c}{\textbf{}} & \multicolumn{1}{c}{\textbf{}} & \multicolumn{3}{c}{\textbf{RILM}}                          & \multicolumn{3}{c}{\textbf{MELS}}                          & \multicolumn{3}{c}{\textbf{SPLSME}}                          \\ \hline
\textbf{Parameter}            & \textbf{Value}           & \textbf{Estimate} & \textbf{Bias} & \textbf{Coverage} & \textbf{Estimate} & \textbf{Bias} & \textbf{Coverage} & \textbf{Estimate} & \textbf{Bias} & \textbf{Coverage} \\ \hline
$\beta^{(2)}_0$                    & -2.20 & -3.21 & -1.01 & 93 & -2.28 & -0.08 & 100 & -2.00 & 0.20  & 98 \\ \hline

$\beta^{(2)}_1$                         & 2.00  & 1.99  & -0.01 & 79 & 2.00  & 0.00 & 96  & 2.00  & 0.00  & 95 \\ \hline

$\beta^{(2)}_2$                         & 0.02  & 0.06  & 0.04  & 96 & 0.02  & 0.00 & 100 & 0.00  & -0.02 & 94 \\ \hline

$\alpha_0^{(2)}$                   & 0.00  & -0.73 & -0.73 & 20 & 0.19  & 0.19  & 99  & 0.08  & 0.08  & 99 \\ \hline

$\alpha_1^{(2)}$                        & 0.30  & -     & -     & -  & 0.33  & 0.03  & 93  & 0.32  & 0.02  & 96 \\ \hline

$\alpha_2^{(2)}$                         & -0.10 & -     & -     & -  & -0.11 & -0.01 & 100 & -0.10 & -0.00 & 99 \\ \hline

$\gamma^{(2)}$                         & -1.00 & -     & -     & -  & -     & -     & -   & -1.32 & -0.32 & 91 \\ \hline

$\delta^{(2)}$                         & 0.50  & -     & -     & -  & -     & -     & -   & 0.68  & 0.18  & 94 \\ \hline

$\sigma_{v_{1, i}^{(2)}}$           & 1.41  & 1.34  & -0.08 & 93 & 1.48  & 0.06  & 98  & 1.53     & 0.12     & 97  \\ \hline

$\sigma_{v_{2, i}^{(2)}}$         & 1.12  & -     & -     & -  & 1.20  & 0.08  & 96  & 1.27     & 0.16     & 93 \\ \hline

$\rho_{v_{1, i}^{(2)}, v_{2, i}^{(2)}}$      & -0.44 & -     & -     & -  & -0.37 & 0.07  & 96  & -0.39     & 0.06     & 96  \\ \hline

$\sigma_{\eta_{1, i}^{(2)}}$              & 1.00  & -     & -     & -  & -     & -     & -   & 0.99  & -0.01 & 94 \\ \hline

$\sigma_{\eta_{2, i}^{(2)}}$            & 1.00  & -     & -     & -  & -     & -     & -   & 1.05  & 0.05  & 93 \\ \hline

$\rho_{\eta_{1, i}^{(2)}, \eta_{2, i}^{(2)}}$         & -0.20 & -     & -     & -  & -     & -     & -   & -0.12 & 0.08  & 98 \\ \hline

$\sigma_{\lambda_i}$              & 1.00  & -     & -     & -  & -     & -     & -   & 1.27  & 0.27 & 95 \\ \hline

$\tau_1$                      & 0.10  & -     & -     & -  & -     & -     & -   & 0.11  & 0.01  & 93 \\ \hline

$\tau_2$                    & 0.05  & -     & -     & -  & -     & -     & -   & 0.00  & -0.05 & 95 \\ \hline

$\tau_3$                    & 0.10  & -     & -     & -  & -     & -     & -   & 0.10  & 0.00  & 99 \\ \hline
\end{tabular}

%% file: bias_0_1_-0.2_1_1_-1.5_0.5_0.1_0.05_0.1.tex
\caption{Bias and coverage rates with $\gamma^{(2)} = -1.50$ and $\delta^{(2)} = 0.50$}
\label{bias_0_1_-0.2_1_1_-1.5_0.5_0.1_0.05_0.1}
\begin{tabular}{lllllllllll}
\hline
\multicolumn{1}{c}{\textbf{}} & \multicolumn{1}{c}{\textbf{}} & \multicolumn{3}{c}{\textbf{RILM}}                          & \multicolumn{3}{c}{\textbf{MELS}}                          & \multicolumn{3}{c}{\textbf{SPLSME}}                          \\ \hline
\textbf{Parameter}            & \textbf{Value}           & \textbf{Estimate} & \textbf{Bias} & \textbf{Coverage} & \textbf{Estimate} & \textbf{Bias} & \textbf{Coverage} & \textbf{Estimate} & \textbf{Bias} & \textbf{Coverage} \\ \hline
$\beta^{(2)}_0$                    & -2.20 & -3.59 & -1.39 & 96 & -2.27 & -0.07 & 100 & -1.96 & 0.24  & 99  \\ \hline

$\beta^{(2)}_1$                         & 2.00  & 1.99  & -0.01 & 86 & 1.99  & -0.01 & 94  & 2.00  & -0.00 & 96  \\ \hline

$\beta^{(2)}_2$                         & 0.02  & 0.08  & 0.06  & 97 & 0.01  & -0.01 & 100 & -0.00 & -0.02 & 96  \\ \hline

$\alpha_0^{(2)}$                   & 0.00  & -0.81 & -0.81 & 14 & 0.07  & 0.07  & 100 & 0.00  & 0.00  & 100 \\ \hline

$\alpha_1^{(2)}$                        & 0.30  & -     & -     & -  & 0.33  & 0.03  & 97  & 0.32  & 0.02  & 97  \\ \hline

$\alpha_2^{(2)}$                         & -0.10 & -     & -     & -  & -0.10 & -0.00 & 100 & -0.10 & 0.00  & 100 \\ \hline

$\gamma^{(2)}$                         & -1.50 & -     & -     & -  & -     & -     & -   & -1.90 & -0.40 & 90  \\ \hline

$\delta^{(2)}$                         & 0.50  & -     & -     & -  & -     & -     & -   & 0.54  & 0.04  & 96  \\ \hline

$\sigma_{v_{1, i}^{(2)}}$           & 1.80  & 1.69  & -0.11 & 95 & 1.87  & 0.06  & 99  & 1.90     & 0.10     & 99   \\ \hline

$\sigma_{v_{2, i}^{(2)}}$         & 1.12  & -     & -     & -  & 1.17  & 0.05  & 97  & 1.24     & 0.12     & 96   \\ \hline

$\rho_{v_{1, i}^{(2)}, v_{2, i}^{(2)}}$      & -0.47 & -     & -     & -  & -0.37 & 0.11  & 96  & -0.38     & 0.09     & 97   \\ \hline

$\sigma_{\eta_{1, i}^{(2)}}$              & 1.00  & -     & -     & -  & -     & -     & -   & 0.95  & -0.05 & 97  \\ \hline

$\sigma_{\eta_{2, i}^{(2)}}$            & 1.00  & -     & -     & -  & -     & -     & -   & 1.05  & 0.05  & 95  \\ \hline

$\rho_{\eta_{1, i}^{(2)}, \eta_{2, i}^{(2)}}$         & -0.20 & -     & -     & -  & -     & -     & -   & -0.13 & 0.07  & 99  \\ \hline

$\sigma_{\lambda_i}$              & 1.00  & -     & -     & -  & -     & -     & -   & 1.24  & 0.24 & 97  \\ \hline

$\tau_1$                      & 0.10  & -     & -     & -  & -     & -     & -   & 0.11  & 0.01  & 93  \\ \hline

$\tau_2$                    & 0.05  & -     & -     & -  & -     & -     & -   & 0.00  & -0.05 & 94  \\ \hline

$\tau_3$                    & 0.10  & -     & -     & -  & -     & -     & -   & 0.10  & -0.00 & 99   \\ \hline
\end{tabular}

%% file: bias_0_1_-0.2_1_1_-0.5_0.75_0.1_0.05_0.1.tex
\caption{Bias and coverage rates with $\gamma^{(2)} = -0.50$ and $\delta^{(2)} = 0.75$}
\label{bias_0_1_-0.2_1_1_-0.5_0.75_0.1_0.05_0.1}
\begin{tabular}{lllllllllll}
\hline
\multicolumn{1}{c}{\textbf{}} & \multicolumn{1}{c}{\textbf{}} & \multicolumn{3}{c}{\textbf{RILM}}                          & \multicolumn{3}{c}{\textbf{MELS}}                          & \multicolumn{3}{c}{\textbf{SPLSME}}                          \\ \hline
\textbf{Parameter}            & \textbf{Value}           & \textbf{Estimate} & \textbf{Bias} & \textbf{Coverage} & \textbf{Estimate} & \textbf{Bias} & \textbf{Coverage} & \textbf{Estimate} & \textbf{Bias} & \textbf{Coverage} \\ \hline
$\beta^{(2)}_0$                    & -2.20 & -2.65 & -0.45 & 92 & -2.04 & 0.16  & 99  & -1.85 & 0.35  & 95  \\ \hline

$\beta^{(2)}_1$                         & 2.00  & 1.98  & -0.02 & 78 & 2.00  & -0.00 & 94  & 2.00  & -0.00 & 97  \\ \hline

$\beta^{(2)}_2$                         & 0.02  & 0.04  & 0.02  & 90 & 0.01  & -0.01 & 98  & 0.00  & -0.02 & 96  \\ \hline

$\alpha_0^{(2)}$                   & 0.00  & -0.70 & -0.70 & 17 & 0.20  & 0.20  & 100 & -0.01 & -0.01 & 100 \\ \hline

$\alpha_1^{(2)}$                        & 0.30  & -     & -     & -  & -0.11 & -0.01 & 100 & -0.09 & 0.01  & 100 \\ \hline

$\alpha_2^{(2)}$                         & -0.10 & -     & -     & -  & 0.34  & 0.04  & 96  & 0.30  & 0.00  & 96  \\ \hline

$\gamma^{(2)}$                         & -0.50 & -     & -     & -  & -     & -     & -   & -0.68 & -0.18 & 96  \\ \hline

$\delta^{(2)}$                         & 0.75  & -     & -     & -  & -     & -     & -   & 0.98  & 0.23  & 90  \\ \hline

$\sigma_{v_{1, i}^{(2)}}$           & 1.12  & 1.07  & -0.05 & 91 & 1.18  & 0.06  & 94  & 1.23  & 0.12  & 94  \\ \hline

$\sigma_{v_{2, i}^{(2)}}$         & 1.25  & -     & -     & -  & 1.29  & 0.04  & 92  & 1.36  & 0.11  & 93  \\ \hline

$\rho_{v_{1, i}^{(2)}, v_{2, i}^{(2)}}$      & -0.41 & -     & -     & -  & -0.29 & 0.13  & 95  & -0.30 & 0.11  & 97  \\ \hline

$\sigma_{\eta_{1, i}^{(2)}}$              & 1.00  & -     & -     & -  & -     & -     & -   & 1.03  & 0.03  & 96  \\ \hline

$\sigma_{\eta_{2, i}^{(2)}}$            & 1.00  & -     & -     & -  & -     & -     & -   & 0.98  & -0.02 & 89  \\ \hline

$\rho_{\eta_{1, i}^{(2)}, \eta_{2, i}^{(2)}}$         & -0.20 & -     & -     & -  & -     & -     & -   & -0.04 & 0.16  & 96  \\ \hline

$\sigma_{\lambda_i}$              & 1.00  & -     & -     & -  & -     & -     & -   & 1.36  & 0.36  & 95  \\ \hline

$\tau_1$                      & 0.10  & -     & -     & -  & -     & -     & -   & 0.11  & 0.01  & 94  \\ \hline

$\tau_2$                    & 0.05  & -     & -     & -  & -     & -     & -   & 0.00  & -0.05 & 95  \\ \hline

$\tau_3$                    & 0.10  & -     & -     & -  & -     & -     & -   & 0.10  & -0.00 & 99  \\ \hline
\end{tabular}